\documentclass[prd,superscriptaddress,floatfix,notitlepage,nofootinbib,reprint]{revtex4-1} % reprint,nofootinbib

\pdfoutput=1
\usepackage{amssymb,graphicx}
\usepackage{amsmath}
\usepackage[]{empheq}
\usepackage{hyperref}
\usepackage{natbib,ifthen}

\usepackage{wasysym}
\usepackage{mathrsfs}
\usepackage[lofdepth,lotdepth,caption=false]{subfig}
\usepackage{color}
\usepackage{bbold}

\newcommand{\commentOut}[1]{}

\newcommand{\be}{\begin{equation}}
\newcommand{\ee}{\end{equation}}
\newcommand{\beq}{\begin{eqnarray}}
\newcommand{\eeq}{\end{eqnarray}}

\begin{document}
%\linenumbers

\newcommand{\jcap}{JCAP}
\newcommand{\apjl}{APJL~}
\newcommand{\aap}{Astronomy \& Astrophysics}
\newcommand{\mnras}{Mon.\ Not.\ R.\ Astron.\ Soc.}
\newcommand{\apjs}{Astrophys.\ J.\ Supp.}
\newcommand{\solphys}{Sol.\ Phys.}
\newcommand{\araa}{ARA \& A}
\newcommand{\pasa}{Publications of the Astronomical Society of Australia}
\newcommand{\jgr}{Journal of Geophysical Research}

\title{Fast Radio Bursts  and the Axion Quark Nugget Dark Matter Model}

\author{Ludovic Van Waerbeke}
\email{waerbeke@phas.ubc.ca}
\author{Ariel  Zhitnitsky}
\email{arz@physics.ubc.ca}
\affiliation{Department of Physics and Astronomy, University of British Columbia, Vancouver, V6T 1Z1, BC, Canada}

\begin{abstract}
 We explore the possibility that the Fast Radio Bursts (FRBs)  are powered  by magnetic reconnection in magnetars, triggered by Axion Quark Nugget (AQN) dark matter.
%The effect, in all respects, is similar to the idea proposed by \cite{Zhitnitsky:2017rop}   for the origin of solar nanoflares. The ``only" distinction is that all physical parameters are drastically (many orders of magnitude) different between these two   systems: the Sun and the magnetar. 
In this model, the magnetic reconnection is ignited by the shock wave which develops when the nuggets' Mach number $M \gg 1$. These shock waves generate very strong  and very short impulses expressed in terms of  pressure $\Delta p/p\sim M^2$  and temperature $\Delta T/T\sim M^2$   in the vicinity of   (would be) magnetic reconnection area. We find that the proposed mechanism produces a coherent emission which is consistent with current data, in particular the FRB energy requirements, the observed energy distribution, the frequency range and the burst duration. Our model allows us to propose additional tests which future data will be able to challenge.

%\vspace{1mm}
\end{abstract} 

\maketitle

% ------------------------------------------------------------------
\section{Introduction}\label{sec:introduction}

Fast Radio Bursts (FRBs) are bright milliseconds duration radio bursts, characterized by Janky-level flux densities and high dispersion measures well above the expected Milky Way contributions \cite{Lorimer-2007,Thornton-2013}. These high dispersion measures, along with many other observational evidence, suggest that they are at cosmological distance (see \cite{Katz-2016} for a review). So far, only one FRB was found to be repeating (FRB121102, see \cite{Spitler-2014}), and recent multi-observatory observations unambiguously associated it with a star forming dwarf galaxy at redshift z=0.193 with a precision of 0.1" \cite{Chatterjee-2017,Tendulkar-2017,Marcote-2017}, therefore there is little doubt that FRBs are cosmological, and we will assume this to be the case throughout this work. Using data from the repeating FRB121102, \cite{Law:2017} showed that the energy distribution $dN/dE_{\rm iso}$ is given by:
\be
\label{FRB_scaling}
\frac{dN}{dE_{\rm iso}}\sim E_{\rm iso}^{-1.7}, ~~~~~~  E_{\rm iso}\in (10^{37}, 10^{40})~ {\rm erg},
\ee
According to \cite{Law:2017}, this slope is indicative of a fundamental underlying physical process, because it is seen by the multi-frequency campaigns performed on FRB121102 using VLA, GBT and Arecibo, even though the burst rate varies by order of magnitudes between observing campaigns. All other known FRBs are single bursts, and whether there is any similarities in the physical processes powering them, and the repeating one, is not established.

Despite the unknown nature of FRBs, their likely cosmological origin implies radiated isotropic energies at the level of $\sim 10^{40}$ erg, corresponding to a very high brightness temperature $> 10^{35}$ K, well above the Compton limit of $10^{12}$ K. Consequently, it is likely that the emission mechanism involves some sort of coherent radiation from "bunches" or "patches" of accelerated charges. Many progenitor models have been proposed, e.g. \cite{popov-2013,Kulkarni-2014,Fuller-2015,Connor-2016,Cordes-2016,Popov-2016} to cite a few, all cosmological solutions invoke a coherent radiation mechanism. The detailed physical processes by which the coherent emission takes place was discussed in \cite{kumar_1, kumar_2}. Their conclusion is that the antenna curvature mechanism is the favored process. A strong magnetic field ($B > 10^{14} ~ G$), typical of magnetars, is necessary, because only then, the electrons in bunches are forced to move collectively along the magnetic field lines, confined in the ground state Landau level and emit coherently for a short period of time.

Current data on FRBs involve energetics, frequency signature, coherence, burst duration and repetition (for FRB121102) and polarization. Proposed models succeed in explaining a subset of constraints, but not the collective data. This situation suggests that FRBs are either have a multi-sources origin or that they represent a new type of phenomena.

In this paper, we explore the possibility that an alternative model of Dark Matter, the Axion Quark Nuggets (AQNs), originally suggested in \cite{Zhitnitsky:2002qa}, could be at the origin of FRBs. AQNs are composite objects of standard quarks in a novel phase, an idea which goes back to quark nuggets  \cite{Witten:1984rs}, strangelets \cite{Farhi:1984qu}, and nuclearities \cite{DeRujula:1984axn} (see also review \cite{Madsen:1998uh} which has a large number of references on the original results).  In the early models \citep{Witten:1984rs,Farhi:1984qu,DeRujula:1984axn,Madsen:1998uh}  the presence of strange quarks stabilizes the quark matter at sufficiently high densities, allowing strangelets being formed in the early universe to remain stable over cosmological timescales. Most of the original models were found to be inconsistent with some observations, but the AQN model was built on different ideas, involving the Axion field, and has not been ruled out so far. In this paper, we quantify the consequences of AQNs falling on a highly magnetized neutron star, and find that it would exhibits an electromagnetic signature similar to the known properties of FRBs. The idea that FRBs could be explained by dark matter falling on a neutron star has been explored by \cite{2017ApJ...844..162T}, but in a context where dark matter are magnetic dipoles that originate from symmetry breaking at the grand unification energy scale. 

To some extend, the physics in our proposal is very similar in spirit to the recent studies \cite{Zhitnitsky:2017rop,Zhitnitsky:2018mav,Raza:2018gpb}, where the authors have proposed that AQNs falling on the Sun could contribute to the solar corona heating and act as the triggers igniting the magnetic reconnections  which generate the  giant solar flares. In the case of FRBs, the central object is a neutron star, and all physical parameters are drastically (many orders of magnitude) different than for the Sun. However, there are two elements   the giant solar flares and FRBs have in common in our framework. First of all, in both cases these events are triggered by AQNs. Secondly,   in both cases the released energy has its source in the magnetic reconnection. Only the environment is vastly different in these two cases.  However, the common origin of these two   phenomena in our framework leads to unambigous prediction that the exponents in the energy distribution for FRBs (\ref{FRB_scaling}) and for solar flares must be numerically very similar.   We will show that this is indeed the case.

The paper is organized as follows.
In Section 2 we review the AQN dark matter model, in Section 3 we discuss the aspects of AQNs going through the Solar corona and flare triggering which are relevant to this study. In Section 4 we develop our proposal following the magnetic reconnection theory for neutron star environment. Lot of the theoretical development has already been suggested in \cite{kumar_1,kumar_2}. Section 5 we compare the results of our proposal with existing data before concluding in Section 6.

%-------------------------------------------------------------------
\section{Axion Quark Nugget (AQN) dark matter model}\label{sec:QNDM}

AQN model stands for the axion quark nugget   model, see the original work \cite{Zhitnitsky:2002qa}  and a short overview
\cite{Lawson:2013bya} with a large number of references on the original results reflecting different aspects of the model.   The  AQN  model   is drastically different from previous similar proposals in two key aspects:\\
1. There is an  additional stabilization factor in the AQN  model provided    by the so-called $N=1$ {\it  axion domain walls}
  which are copiously produced during the QCD transition.
   \\
  2. The AQN  could be
made of matter as well as {\it antimatter} in this framework as a result of separation of charges, see  recent papers \cite{Liang:2016tqc, Ge:2017ttc,Ge:2017idw} for technical details.

 The most  important implication for the present studies
  is that quark nuggets made of  antimatter
 store a huge amount of energy that can be released when the anti-nuggets from outer space hit a star and are annihilated in the star's atmosphere.  This feature
 of the AQN model is unique because, unlike any other dark matter models, AQNs are made  of the same quarks and antiquarks of the standard model (SM) of particle physics. One should also remark that the annihilation of anti-nuggets with visible matter  may  produce a number of other observable effects such as  rare events of annihilation of anti-nuggets with  visible matter     in the centre of galaxy, or in the    Earth atmosphere,  see  references on the original computations in \cite{Lawson:2013bya}   and a few comments at the end of this   section.

The basic idea of  the AQN  proposal can be explained   as follows:
It is commonly  assumed that the Universe
began in a symmetric state with zero global baryonic charge
and later (through some baryon number violating process, the so-called baryogenesis)
evolved into a state with a net positive baryon number. As an
alternative to this scenario we advocate a model in which
``baryogenesis'' is actually a charge separation process
when  the global baryon number of the Universe remains
zero. In this model the unobserved antibaryons come to comprise
the dark matter in the form of dense nuggets of quarks and antiquarks in colour superconducting (CS) phase.
  The formation of the  nuggets made of
matter and antimatter occurs through the dynamics of shrinking axion domain walls, see
original papers \cite{Liang:2016tqc,Ge:2017ttc} for technical  details.

 The  nuggets, after they form,  can be viewed as strongly interacting and macroscopically large objects with a  typical  nuclear density
and with a typical size $R\sim (10^{-5}-10^{-4})$cm determined by the axion mass $m_a$ as these two parameters are linked, $R\sim m_a^{-1}$.
It is important to emphasize that there are strong constraints on the    allowed window for the axion mass: $10^{-6} {\rm eV}\leq m_a \leq 10^{-2} {\rm eV}$, see original papers \cite{axion1,axion2,axion3,KSVZ1,KSVZ2,DFSZ1,DFSZ2} and   reviews \cite{vanBibber:2006rb,Asztalos:2006kz,Sikivie:2008,Raffelt:2006cw,Sikivie:2009fv,Rosenberg:2015kxa,Graham:2015ouw,Ringwald:2016yge}  on the theory of the axion and recent progress on axion search experiments.

 This axion window corresponds to the range of the nugget's baryon charge $B$ which   largely overlaps  with all available and independent constraints:
\beq
 \label{B-range}
 10^{23}\leq |B|\leq 10^{28},
 \eeq
 see e.g. \cite{Jacobs:2014yca,Lawson:2013bya} for reviews.
  The corresponding mass $\cal{M}$ of the nuggets  can be estimated as ${\cal{M}}\sim m_pB$, where $m_p$ is the proton mass.
The lower bound comes from the fact that axions with mass less than $10^{-6} {\rm eV}$ would produce too much dark matter and conflict with cosmological constraints.
The upper bound comes from astrophysical constraints: if axions are more massive than $10^{-2} {\rm eV}$, radiative cooling of white dwarfs and supernova SN1987A would
exceed observations. 

We should comment here that the axion's   contribution to the dark matter density scales as $\Omega_{\rm axion}\sim m_a^{-7/6}$. This scaling  unambiguously implies that the axion mass must be fine-tuned  $m_a\simeq 10^{-6} $ eV
 to  saturate the DM density today  while larger axion mass will contribute very little to $\Omega_{\rm DM}$.
 In contrast, the AQN's   contribution  to $\Omega_{\rm DM}$ is not sensitive to the axion mass $m_a$ and always satisfies the condition (\ref{Omega}). It is expected that all three distinct production mechanisms (misalignment mechanism, topological defect decays and  the AQN production) will be operational  during the QCD transition as discussed in details in 
 \cite{Ge:2017idw}. However, only the AQN production mechanism will always satisfies (\ref{Omega}) irrespectively to the axion mass, while other mechanisms will be subdominant for sufficiently large axion mass $m_a\geq 10^{-4}$ eV. 

We should note that within the mass range (\ref{B-range}), direct detection of the AQNs  is not possible with conventional  particle detectors
designed for WIMP (Weakly interacting Massive Particle) searches because   the average impact rate is approximately
one AQN per year per ${\rm km}^2$.   At the same time, the   detectors with a large area designed for analyzing of the high energy cosmic rays, such as Pierre Auger Observatory and Telescope Array in principle, are   capable   to study AQNs, see \cite{Lawson:2013bya} for short overview.

 The AQN model is perfectly consistent with all known astrophysical, cosmological, satellite and ground based constraints within the parametrical range for
 the mass ${\cal M}$ and the baryon charge $B$ mentioned   above (\ref{B-range}). It is also consistent with known constraints from axion search experiments. Furthermore, there is a number of frequency bands where an excess of emission was observed, but not explained by conventional astrophysical sources. The AQN model explains some portion, and may even explain the entire excess of radiation in these frequency bands, see short review \cite{Lawson:2013bya} and additional references at the end of this section.

Another key element of the AQN model is that the coherent axion field $\theta$ is assumed to be non-zero during the QCD transition in early Universe.
       As a result of these $\cal CP$ violating processes, the number of nuggets and anti-nuggets
      being formed is different. This difference is an order of one effect   \cite{Liang:2016tqc,Ge:2017ttc}, irrespective to the axion mass $m_a$ or initial angle $\theta_0$.  
      %In contrast, in conventional cosmology, baryogenesis requires extreme fine tuning to explain the small matter excess over anti-matter. 
      As a result of this disparity between nuggets and anti nuggets, a similar disparity would also emerge between visible quarks and antiquarks, which would survive until present day.
       This  is precisely  the reason why the resulting visible and dark matter
densities, while being different, have the same order of magnitude \cite{Liang:2016tqc,Ge:2017ttc}
\be
\label{Omega}
 \Omega_{\rm dark}\sim \Omega_{\rm visible}
\ee
as they are both proportional to the same fundamental $\Lambda_{\rm QCD} $ scale,
and they both originate at the same  QCD epoch.
  If these processes
are not fundamentally related, the two components
$\Omega_{\rm dark}$ and $\Omega_{\rm visible}$  could easily
exist at vastly different scales.

  Unlike conventional dark matter candidates, such as WIMPs
 the matter and antimatter
nuggets are strongly interacting macroscopically large objects.
However, they do not contradict any of the many known observational
constraints for cold dark matter or
antimatter    in the Universe due to the following reasons~\cite{Zhitnitsky:2006vt}:
  They carry  very large baryon charge
$|B|  \gtrsim 10^{23}$, so their number density is very small ($\sim B^{-1}$).
 As a result of this unique feature, their interaction  with visible matter is highly  inefficient, and
therefore, the AQNs perfectly qualify  as  Cold Dark Matter  candidates. Furthermore,
  the AQNs have  very  large binding energy due to the   large    gap $\Delta \sim 100$ MeV in  CS phases.
Therefore, the baryon charge is so strongly bounded in the core of the nugget that  it  is not available to participate in big bang nucleosynthesis
(\textsc{bbn})  at ${\rm k_B} T \approx 1$~MeV, long after the nuggets were formed. Therefore, it does not modify conventional \textsc{bbn} physics. In fact, it may help to resolve  the so-called  ``Primordial Lithium Puzzle", see
below.  

 It should be noted that the galactic spectrum
contains several excesses of diffuse emission of unknown origin, the best
known example being the strong galactic 511~keV line.
%\exclude{If the nuggets have the  average  baryon
%number in the $\langle B\rangle \sim 10^{25}$ range they could offer a
%potential explanation for several of
%these diffuse components.
%(including 511 keV line and accompanied   continuum of $\gamma$ rays in 100 keV and few  MeV ranges,
%as well as x-rays,  and radio frequency bands).
%It is important to emphasize that a comparison between   emissions with drastically different frequencies in such  computations
% is possible because the rate of annihilation events (between visible matter and antimatter DM nuggets) is proportional to
%one and the same product    of the local visible and DM distributions at the annihilation site.
%The observed fluxes for different emissions thus depend through one and the same line-of-sight integral
%\be
%\label{flux1}
%\Phi \sim R^2\int d\Omega dl [n_{\rm visible}(l)\cdot n_{DM}(l)],
%\ee
%where $R\sim B^{1/3}$ is a typical size of the nugget which determines the effective cross section of interaction between DM and visible matter. As $n_{DM}\sim B^{-1}$ the effective interaction is strongly suppressed $\sim B^{-1/3}$. }
The parameter $\langle B\rangle\sim 10^{25}$  was fixed in this  proposal by assuming that this mechanism  saturates the observed  511 keV line   \cite{Oaknin:2004mn, Zhitnitsky:2006tu}, which resulted from annihilation of the electrons from visible matter and positrons from anti-nuggets.   Other emissions from different frequency bands  are expressed in terms of the same parameter $\langle B\rangle$, and therefore, the  relative  intensities  are unambiguously and completely determined by the internal structure of the AQNs which is described by conventional nuclear physics and basic QED, see
\cite{Lawson:2013bya} for a short overview with references on specific computations of diffuse galactic radiation  in different frequency bands.

Another place where AQN model may lead to some observable effects is 
study of possible extra emission which may occur at earlier times and may be observable today.  
In fact,  the recent EDGES observation of a stronger than
anticipated 21 cm absorption  \cite{Bowman:2018yin} can naturally find its  explanation within the AQN framework as recently advocated in
\cite{Lawson:2018qkc}.  The basic idea is that the extra   thermal
emission from AQN  dark matter at   $z\simeq 17$   produces the required intensity (without  adjusting of any parameters) to explain the recent EDGES observation.

Last, but not least. 
 The AQN  model may offer a very natural resolution of the so-called ``Primordial Lithium Puzzle" as recently argued in 
\cite{Flambaum:2018ohm}. This problem  has been with us for at least two decades, and   conventional astrophysical and nuclear physics  proposals could not resolve this longstanding mystery. In the AQN framework this puzzle is automatically and naturally resolved without adjusting any parameters assuming the same window (\ref{B-range}) for the baryon charge of the nuggets, as argued  in \cite{Flambaum:2018ohm}. This resolution represents yet another, though indirect, support for this new AQN framework.

%-------------------------------------------------------------------
\section{AQNs as solar flares triggers}\label{AQN-flares}

In this section we  overview the basic results from \cite{Zhitnitsky:2018mav} which argued that the AQNs play the role of {\it triggers} for Solar flares. We want to avoid a possible confusion between the large flares when the AQNs play an auxiliary  role sparking the large (but rare) events in form of  flares and the AQNs when they play the principle  role of the heating corona with emission of the EUV radiation  as studied in 
 \cite{Zhitnitsky:2017rop,Raza:2018gpb}. In former case the dominant portion of the energy comes from magnetic field 
 in active regions of the Sun, while in later case the energy for each event (which are identified with solar nanoflares) 
 is a result of annihilation of the AQNs in corona.  The corresponding events (nanoflares) are uniformly distributed over the solar surface and with the frequency of appearance which depends very modestly   on a specific time during a  solar  cycle. It should be contrasted with large solar flares which are very rare energetic events, and which are originated exclusively in active regions with large magnetic field. The corresponding frequency of appearance may vary by factor of 100 during the solar cycle. 
 
It turns out that if one computes    the extra energy being produced within the AQN dark matter scenario,     one obtains the total extra energy due to the AQN annihilation events on the level $\sim 10^{27}{\rm erg}/{\rm  s}$    which  reproduces     the   observed EUV and soft x-ray intensities  \cite{Zhitnitsky:2017rop,Raza:2018gpb}. One should add that the estimate   $\sim 10^{27}{\rm erg}/{\rm  s}$  for extra energy  is derived  exclusively in terms of the known  dark matter density $\rho_{\rm DM} \sim 0.3~ {\rm GeV\cdot cm^{-3}}$ surrounding the sun without adjusting any  parameters of the model.

The AQNs entering the Solar Corona activate the magnetic reconnection of {\it preexisting}  magnetic configurations in    active regions (distributed very non-uniformly). Technically the effect  occurs  due to the {\it shock waves} which form as the typical velocity of the   nuggets $v\sim (600-800)~ {\rm km/s}$ in the vicinity of the surface, is well above the speed of sound, $v/c_s\gg 1$. The energy of the flares in this case is determined  by the  preexisted magnetic field configurations occupying very large area in active region, while the relatively small amount of energy associated with initial AQNs plays a minor role in the total  energy released during a large flare.
First, we  list  the  key elements     of this framework which will be  crucial    in our proposal on the nature of FRBs to be formulated  in section \ref{FRB}.

 1. AQNs entering the solar corona will inevitably  generate shock waves as  the  speed of sound $c_s$ is smaller than
 the typical  velocities $v$ of the dark matter particles (Mach number $M=v/c_s> 1$);

 2. When the AQNs (distributed uniformly) enter regions with a strong magnetic field, they trigger magnetic reconnection of {\it preexisted} magnetic fluxes in {\it active regions};

 3. Technically, the AQNs are capable sparking magnetic reconnections due to the large discontinuities of the pressure $\Delta p/p\sim M^2$ and temperature $\Delta T/T\sim M^2$ when the shock front passes through the magnetic reconnection regions;

 4. The energy of the flares $E_{\rm flare}\sim L^2_{\perp}$ is powered  by the   magnetic field occupying very large area  $\sim L^2_{\perp}$ in active regions;

 5. The above mechanism, with few additional assumptions, predicts that the number of flares $dN(E_{\rm flare})$ with energy of order $E_{\rm flare}$ scales as $dN\sim E_{\rm flare}^{-1}$, consistent  with observations;

 6. This proposal   naturally resolves the problem of drastic separation of scales when a class-A pre-flare (measured by X-rays) lasts  for seconds, a flare itself lasts about an hour,  while the preparation phase of the magnetic configurations (to be reconnected during the flare) may last months. These three  drastically different time-scales can peacefully coexist in our framework because the trigger is not an internal part of the magnetic reconnection's dynamics.

We now elaborate on a few elements from this  list to emphasize the most important fundamental features that will apply to AQNs as triggers for FRBs.
% The corresponding formulae and estimates will provide some orientation for our main studies  in section \ref{FRB} on the FRBs, when basic  principles are applied to a neutron star (NS) with drastically different parameters  (size, magnetic field, density and temperature).
% However, the fundamental   physical principles  remain the same and the role of the solar flares is played by   FRBs in case of a NS.
%The AQNs initiate the magnetic reconnections   in both systems when they enter the active regions with preexisting magnetic configurations characterized by large field gradients.

% The nanoflares, identified with  the AQN annihilation events heat the  corona and
% are the main source of the  EUV radiation (\ref{total_power}). There are strong arguments   which suggest  that the EUV radiation is   correlated with large M, X -flares
% as shown in    \cite{Bertolucci-2017}. Therefore, one should expect that these two phenomena (nanoflares and M,X-flares), characterized by drastically different scales, must originate from the same physics. 
% We overview  this proposal in next subsections by
%emphasizing the most important fundamental features which apply to the Sun and the NS.

  \subsection{AQNs and  shock waves in plasma}\label{shock}

 We start our  overview with  estimations   of the speed of sound $c_s$ in the corona at $T\simeq 10^6 K$,
 \beq
 \label{sound}
 \left(\frac{c_s}{c}\right)^2 &\simeq  &\frac{3\Gamma ~{\rm k_B} T}{m_p c^2},  ~~~ c_s\simeq 2\cdot 10^{-4} c\cdot \sqrt{ \frac{T}{10^5~ {\rm K}}} \\
 c_s &\simeq  &0.6\cdot 10^7 \sqrt{ \frac{T}{10^5~ {\rm K}}}\cdot  \left(\frac{\rm cm}{\rm s} \right), \nonumber
 \eeq
 where $\Gamma=5/3$ is a specific heat ratio, $c$ is the speed of light, ${\rm k_B}$ the Boltzmann constant and we approximate the mass density $\rho_p$ of plasma by the proton's number density density $n$ as follows $\rho_p\simeq n m_p$. The crucial observation here is that the Mach number $M$ is always  much  larger than one for a typical dark matter velocities\footnote{A typical velocity of the nuggets $v_{AQN}$ when they enter the solar atmosphere
 is few times larger than the velocity  $v$ computed far away from the Sun as mentioned in footnote \ref{velocity}.}:
 \be
 \label{Mach}
 M\equiv \frac{v_{AQN}}{c_s}\simeq 10  \sqrt{ \frac{10^5~ {\rm K}}{T}} \gg 1.
 \ee
 As a result, a strong shock waves will be generated when the AQNs enter the solar corona.
  In the limit when the thickness of the shock wave can be ignored,  the corresponding formulae for  the discontinuities of the  pressure $p$, temperature $T$, and the density $\rho_p$ are well known and given by, see e.g. \cite{Landau}:
  \beq
 \label{shock1}
 \frac{{\rho_p}_2}{{\rho_p}_1} &\simeq  &\frac{(\Gamma+1) }{(\Gamma-1)}, ~~~~~  \frac{p_2}{p_1}\simeq M^2\cdot \frac{2\Gamma }{(\Gamma+1)} \nonumber \\
 \frac{T_2}{T_1} &\simeq  &M^2\cdot \frac{2\Gamma(\Gamma-1) }{(\Gamma+1)^2}, ~~~~ M \gg 1,
 \eeq
 where we assume $M\gg 1$ and keep the leading terms only in the corresponding formulae.

 \subsection{Magnetic reconnection ignited by the shock waves}\label{m_reconnection}
%\exclude{
%We define   the dimensionless  parameter $\beta$   determining  the importance of the magnetic pressure in plasma as follows
% \be
% \label{beta}
% \beta\equiv \frac{8\pi  p}{B^2} \sim 10^{-2}\left(\frac{n}{10^{10} ~{\rm cm^{-3}}}\right)\cdot\left(\frac{T}{10^6 K}\right)\cdot\left(\frac{100~ G}{B}\right)^2,
% \ee
% where for numerical estimates we use typical parameters for the active regions in corona when $\beta\ll 1$.
% }
The Alfv\'{e}n velocity  $v_A$   assumes the following typical numerical values in the  corona environment
 \beq
 \label{alfven}
 \frac{v_A}{c}&=&\frac{{\cal{B}}}{c\sqrt{4\pi\rho_p}} \sim \left(\frac{{\cal{B}}}{100~ G}\right)\cdot \sqrt{ \frac{10^{10} ~{\rm cm^{-3}}}{n} } \nonumber \\
& \sim & 2\cdot 10^{-3} \\
 v_A &\simeq & 600  ~ \frac{\rm km}{\rm s}. \nonumber
 \eeq
 One can introduce the Alfv\'{e}n Mach number $M_A$ which plays a role similar to the Mach number:
 \be
 \label{M_A}
 M_A\equiv \frac{v_{\rm AQN}}{v_A}.
 \ee
 Numerically, $M_A$ could be larger or smaller than unity depending on the numerical values of the density $n$ and the magnetic field  ${\cal{B}}$.  In what follows we assume that some finite fraction of the nuggets will have $M_A > 1$, similar to our assumption that $M\gg 1$.   In this case one should expect that  very  fast shock   develops which may trigger the large flare.

The idea   that the shock waves may drastically increase the rate of magnetic reconnection is not new, and has been discussed previously in the literature \cite{Tanuma}, though a quite different context: it was applied to interstellar medium in the presence of a supernova shock. \cite{Tanuma} showed that the shock waves may trigger and ignite     sufficiently fast reconnections\footnote{\label{MHD}Furthermore, the 2d MHD simulations \cite{Tanuma} show  that  a large number of different phenomena,
    including SP reconnection \cite{Sweet, Parker1}, Petschek reconnection \cite{Petschek, Kulsrud}, tearing instability, formation of the magnetic islands, and many others, may all take place at different phases in  the evolution of the system, see also reviews \cite{Shibata:2016,Loureiro}.}. The new element which was  advocated in
    \cite{Zhitnitsky:2018mav}
   is that the small shock waves  resulting from  entering AQNs are widespread and generic events in the solar corona. If the nuggets enter the active regions of strong magnetic field,  they may ignite large flares due to magnetic reconnections.    This is precisely the correlation which has   been observed    in \cite{Bertolucci-2017}, and which was the main motivation for proposal \cite{Zhitnitsky:2018mav}.

  The key elements suggesting that the fast shock  wave may drastically modify  the rate of magnetic reconnection
 is based on observation that the pressure,  the temperature and the magnetic pressure  may   experience some dramatic changes when the shock wave approaches the reconnection region. To be more precise, the  formulae (\ref{shock1}) must be modified  in the presence  of  a magnetic field  \cite{Landau}:  One should  replace $p_i\rightarrow p_i^*$, where $p_i^*\equiv p_i+\frac{{\cal{B}}_i^2}{8\pi}$ to account for the magnetic field pressure.

   \subsection{\label{scaling}Geometrical interpretation of the scaling $ {dN} \sim   E^{-\alpha} dE$ }

\begin{figure*}
\centering
\includegraphics[width=0.8\textwidth]{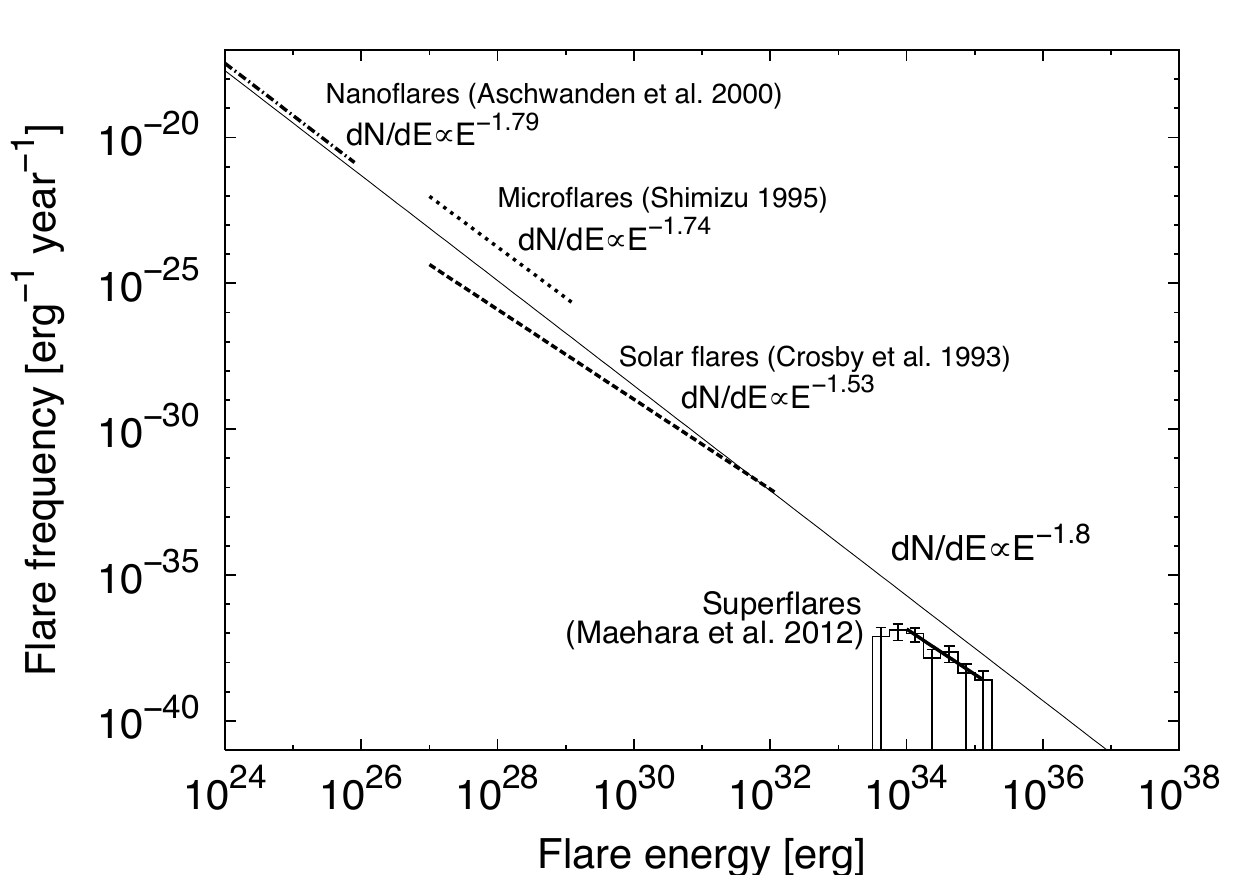}
\caption{\label{Shibata:2016} Energy distributions for all types of solar flares; microflares, nanoflares and superflares.  The plot is  adopted  from \citep{Shibata:2016}.}
\end{figure*}

 In this subsection we overview the arguments presented  in \cite{Zhitnitsky:2018mav}
 that  the observed scaling $dN\sim E^{-\alpha}dE$ of    the flare's frequency appearance   as a function of the  energy $E$
can be interpreted in geometrical terms within the AQN scenario.   The observational  data can be described sufficiently well with a single  exponent  $\alpha\simeq 2$  covering vastly different energy scales covering 12 orders of magnitude,
  including nanoflares, microflares, solar flares, and even superflares, see e.g. Fig \ref{Shibata:2016} adopted from
 review \cite{Shibata:2016}.

It is convenient to represent the same scaling behaviour by  integrating   formula $dN\sim E^{-\alpha}dE$    over energy $dE$ (to be more specific, let us  say from $E$ to $2E$)   to represent it in the following form
 \be
  \label{distribution2}
   {N} (E) \simeq   C\left(\frac{E_0}{E}\right) ~~~~~~ {\rm for} ~~~~ \alpha\simeq 2 ~~ {\rm and } ~~E\simeq  (  10^{26}  -  10^{32})~{\rm erg},
  \ee
 where we keep the  energetic window corresponding to   microflares and flares.  The coefficient $C\sim 10^4 ~{\rm flares/year}$ is a normalization factor that can be determined from the observations. The dimensional parameter  $E_0\simeq 10^{26} ~{\rm erg}$ is  a   part of this normalization's convention. In these notations the  $N(E)\sim E^{-1}$   represents  the number of flare events per year with energy of order $E$.    Can we interpret this scaling within the AQN framework? The answer is yes, and the argument goes as follows.

The probability that a  shock wave cone with size $\sim R^2$ passes through a much larger sunspot area $L_{\perp}^2$ is proportional to suppression factor $(R/L_{\perp})^2$. The same sunspot area $L_{\perp}^2$ also enters the expression for the volume $V_{\rm flare}=(L_{\perp}^2L_z)$ in active region,  which describes the  total magnetic  energy    potentially available for   its   transferring  into the flare heating  as a result of magnetic reconnection. Assuming that a typical averaged magnetic field ${\cal B}$ over the entire regions and the  relevant  height $L_z$ of the solar atmosphere
are external parameters with respect to the magnetic reconnection dynamics, one can infer that the energy of the flare  scales   as
 $E_{\rm flare}\sim L_{\perp}^2$, while   the probability to ignite the magnetic reconnection in area  $L_{\perp}^2$ is proportional to $(R/L_{\perp})^2\sim E^{-1}_{\rm flare}$, as discussed above. Therefore,     the observed relation   (\ref{distribution2}) in our framework has a pure geometrical interpretation, as
    the frequency $N$ of appearance of a flare with energy of order $E_{\rm flare}$ scales\footnote{\label{deviation}It is quite  obvious  that there is a number of corrections which may modify our theoretical  prediction (from  $\alpha\simeq 2$ to the observed scaling with $\alpha\simeq 1.8$ shown in Fig.\ref{Shibata:2016}).
    We shall not elaborate on   possible causes  for the deviations in this paper as it is well beyond  the scope  of the present work.} as
$N \sim L_{\perp}^{-2}\sim E^{-1}_{\rm flare}$.

 As these arguments are purely geometrical in nature, and do not depend on specific features of a star,  they can be applied to any system, including a NS,   when the corresponding
 explosion-like flare is sparked by AQNs.  As will be argued in next section, the corresponding event  in a  NS will  be identified with FRBs. The immediate consequence of this identification is that the frequency of appearance  of the FRBs must follow the  same
 scaling $dN\sim E^{-2}dE$ discussed above. As we reviewed in Introduction (see (\ref{FRB_scaling})  this scaling with $\alpha\simeq 1.8$ is consistent with   our oversimplified  geometrical interpretation, and almost identically coincides with data  shown  on Fig.\ref{Shibata:2016}   with $\alpha\simeq 1.8$ representing
 the solar flare's observations.

%-------------------------------------------------------------------
\section{AQNs as the triggers for the Fast Radio Bursts}\label{FRB}

We want to present some arguments suggesting that the observed FRB explosions are a result of magnetic reconnections in NS, similar to conventional views that solar flares are due to the magnetic reconnections in active Sun regions. In both cases the energy of the explosions is powered by existing magnetic fields.  The  idea that FRBs are a result of magnetic reconnection in NS was advocated in \cite{kumar_1, kumar_2}. The new element proposed in the present work is that the FRBs are {\it triggered} by the AQN hitting the NS from outer space, similar to their role in the solar flares.
The moment when the AQN is initiating and activating the magnetic reconnection is the main subject of the present work. Before we present our arguments supporting the role of AQNs, we highlight   in subsection \ref{SP} the conventional viewpoint on the old Sweet-Parker (SP)  theory of the magnetic reconnection,  its main results and its main problems, and explore how AQNs can trigger FRBs, using the analogy with solar flares. We continue in subsection \ref{sect:Mach1} to argue that a shock wave is likely to develops when the AQN enters the NS.  Finally, in subsection \ref{FRB_parameters} we argue that the parameters  which were postulated in \cite{kumar_1, kumar_2} for FRBs naturally emerge  from our framework.

  \subsection{\label{SP}Sweet-Parker's (SP) theory }

 We start by introducing the most important parameters of the problem:
 \beq
 \label{definitions}
 S&=& \frac{Lv_A}{\chi_m}, ~~~~~ v_A=\frac{{\cal B}}{\sqrt{4\pi h}}, ~~~~~ \chi_m=\frac{c^2}{4\pi\sigma} \\
 h&=&\frac{c^2\rho_p+{\cal B}^2/(4\pi)}{c^2}\simeq \frac{{\cal B}^2}{4\pi c^2} \nonumber 
 \eeq
where  $S$ is the so-called the Lundquist number,   $L$ is the typical size of the region which will host the reconnection,   $v_A$ is Alfv\'{e}n velocity, $\rho_p$ is the plasma's mass density,  $\chi_m$ is the magnetic diffusivity, and finally  $\sigma$ is the electrical conductivity of the plasma. We also note that the definition for Alfv\'{e}n velocity  is given  in terms of enthalphy $h$ rather than in conventional way in terms of mass density $\rho_p$. This is because $v_A\sim c$ is close to the speed of light  in NS environment  when the term ${\cal B}^2$ is large, and  cannot be ignored. Such a definition for $v_A$ is a  common practise  in NS physics, dynamics of the relativistic jets, and other areas when  $v_A\sim c$ (see e.g. \cite{rezania}).

The most important parameter for our future estimates is the dimensionless parameter $S$.  For  typical coronal conditions, $S\sim (10^{12}-10^{14})$. For the NS environment the parameter  $S$ could be event larger $S\sim (10^{17}-10^{20})$,
because  $\chi_m$ is many orders of magnitude smaller for NS environment in comparison with the solar environment.  These large numerical values for both systems play an important role in our  arguments related to the speed of the magnetic reconnection and its relation to the SP theory.

The original ideas on magnetic reconnection was formulated   by \cite{Sweet}  and \cite{Parker1}.
Using simple dimensional arguments, Sweet and Parker (SP) have shown that the reconnection time $\tau_{\rm rec}$ is
quite slow and expressed in terms of the original parameters of the system as follows:
\beq
\label{eq:SP}
\frac{\tau_A}{\tau_{\rm rec}} &\sim  &\frac{1}{\sqrt{S}}, ~~~~ \tau_A\equiv\frac{L}{v_A}, ~~~~ \frac{u_{\rm in}}{u_{\rm out}}\sim  \frac{1}{\sqrt{S}} \\
 \frac{l}{L} &\sim  &\frac{1}{\sqrt{S}}, ~~~~~\tau_{\rm rec}\sim \frac{L}{u_{\rm in }} \nonumber 
\eeq
 where $u_{\rm in}$ is the  velocity of reconnection between oppositely directed fluxes of thickness $l$, and  $u_{\rm out}\sim v_A$ is normally assumed to be of   order of  the Alfv\'{e}n velocity. The scaling relations (\ref{eq:SP}) predicted by the SP theory are insufficient to explain the reconnection rates observed in the corona due to the very large numerical values of $S\sim (10^{12}-10^{14})$. The next step to speed up the reconnection rate has been undertaken in \cite{Petschek} with some important amendments in \cite{Kulsrud} where it was  argued that the reconnection rate could be much faster than the  original formula (\ref{eq:SP}) suggests.
 However, some subtleties remained in the proposal \cite{Petschek,Kulsrud}. Furthermore, the numerical simulations reproduce conventional scaling formulae (\ref{eq:SP}) at least for moderately large $S\lesssim 10^4$. In the last 10-15 years
 a large number of new ideas have been pushed forward. It includes such processes as plasmoid- induced reconnection,  fractal reconnection, to name just a  few. It is not the goal of the present work to analyze  the assumptions, justifications, and the problems related to proposals \cite{Sweet,Parker1,Petschek,Kulsrud},  and we refer to
 the recent review papers \cite{Shibata:2016,Loureiro} for recent developments and relevant discussions  on these matters. We are only underlying that AQNs acting as triggers (through shock waves) can drastically speed up the reconnection as     reviewed in previous section, and provide a physical context consistent with the ideas of \cite{kumar_1,kumar_2}. Similarly,  as shown in \cite{Tanuma}, the shock waves may trigger and ignite (in very different circumstances) sufficiently fast reconnections (see footnote \ref{MHD} for references and details).
There are few  important parameters which control  the dynamics of the  system: in addition  to $S$  it is convenient to introduce another dimensionless  parameter $\beta$ which determines the importance of the magnetic pressure in comparison with the gas pressure,
 \be
 \label{beta1}
 \beta\equiv \frac{8\pi  p}{{\cal B}^2} \sim  10^{-14}\left(\frac{n}{10^{18} ~{\rm cm^{-3}}}\right)\cdot\left(\frac{{\rm k_B} T}{1 ~{\rm MeV}}\right)\cdot\left(\frac{10^{14}~ G}{{\cal B}}\right)^2,
 \ee
 where for numerical estimates we use the  parameters for NS environment relevant for FRBs, which are required
 for the observations to be consistent with the idea that FRBs are powered by magnetic reconnections as  discussed in \cite{kumar_1,kumar_2}.
 Specifically,   the number density $n\sim 10^{18} ~{\rm cm^{-3}}$  and typical magnetic field ${\cal B}\sim 10^{14}~G$ have been estimated in  \cite{kumar_1,kumar_2} by analyzing the  observational FRB's properties.  The density $n$ corresponds to the electron number density  in the region where the magnetic reconnection is expected to occur in the NS atmosphere.  The parameter $\beta$ in (\ref{beta1}) is much smaller for the NS environment in comparison with the $\beta\sim 10^{-2}$ in the solar corona (due to drastically stronger magnetic field in NSs). One should emphasize that parameter $\beta$ as estimated in (\ref{beta1}) represents some average global characteristic of the system. Locally, parameter $\beta$ may drastically deviate from (\ref{beta1}) due to strong variation of the magnetic field in vicinity of the magnetic reconnection area, see some comments at the end of section \ref{sect:Mach1}.

  Another important parameter in  our discussions is the speed of sound  $c_s$ in the region of the NS atmosphere
  where the reconnection might occur  at ${\rm k_B} T\sim 1~ {\rm MeV}$ and  $n\sim 10^{18} ~{\rm cm^{-3}}$. The electrons at this temperature are relativistic as $T\sim m_e$. However, the protons remain to be non-relativistic, and we assume that the proton's density, on average,   is the same  order of magnitude as the electron's density $n$.
  With this assumption we estimate the speed of sound, similar to  (\ref{sound}) as follows:
 \be
 \label{sound1}
\left(\frac{c_s}{c}\right)^2\sim \frac{3\Gamma ~{\rm k_B} T}{m_p c^2},  ~~~ c_s\simeq  7\cdot 10^{-2} c\cdot \sqrt{ \frac{{\rm k_B} T}{1~ {\rm MeV}}}.
   \ee
  One should emphasize  that the speed of sound  in the interior of NSs  is different from the expression given by eq. (\ref{sound1}). In ultra-relativistic case (which includes the density colour superconductor phase), it can be explicitly computed $ {c_s}=  (1/\sqrt{3})c$. As we shall discuss below the estimate for  the speed of sound in NS atmosphere (\ref{sound1}) will play an extremely important role in what follows because it shows that it is typically smaller than the velocity of the AQNs entering the NS atmosphere with  $v_{\rm AQN}\simeq c$ (as discussed in next section, see eq. \ref{Mach1}).
 In such circumstances  a high speed nuggets will  generate a shock wave, which is the key element for our arguments.

 \subsection{Mach number and shock waves}\label{sect:Mach1}
 First of all we want to demonstrate  that the typical velocity of the AQNs captured by a NS  will always be close to the speed of light, irrespectively of the initial velocity distribution of the dark matter particles when ``measured"   far away from the NS. Indeed,
  the impact  parameter for  capture and crash of the AQNs by the NS can be estimated as follows,
  \beq
  \label{capture1}
  \frac{b_{\rm cap}}{ R_{\rm NS}}&=&\sqrt{1+\gamma_{\rm NS} }\simeq \frac{c}{v_{\infty}}\sqrt{ \frac{2GM_{\rm NS}}{R_{\rm NS}c^2} } \\
&\simeq &\frac{c}{v_{\infty}}\cdot \left(\frac{\rm 10^6~cm}{R_{\rm NS}}\right)^{1/2} \cdot \left(\frac{M_{\rm NS}}{2~M_{\odot}}\right)^{1/2} , \nonumber 
  \eeq
  where $v_{\infty}\simeq 10^{-3}c$ is a typical velocity of the nuggets at a large distance from the star. From (\ref{capture1}) one can explicitly see that the capture impact parameter $\sim (c/v_{\infty})$ is many orders of magnitude larger than ${ R_{\rm NS}}$. This feature   should be  contrasted with the solar system   
  where ${b_{\rm cap}}$  is only few times larger than the solar radius $R_{\odot}$. One can estimate a typical longitudinal velocity $v_{\parallel}$ (along the NS's surface) of a nugget   from angular momentum conservation as follows,
   \be
  \label{velocity}
  v_{\parallel} R_{\rm NS}\simeq b v_{\infty} ~~\Rightarrow ~~   v_{\parallel} \simeq c\left(\frac{b}{b_{\rm cap}}\right)\cdot \left(\frac{\rm 10^6~cm}{R_{\rm NS}}\right)^{1/2} \cdot \left(\frac{M_{\rm NS}}{2~M_{\odot}}\right)^{1/2}.
     \ee
 A typical velocity component $v_{\perp}$ (perpendicular to the NS surface) can be estimated as a free fall velocity. It assumes the same order of magnitude as (\ref{velocity}), i.e $v_{\perp}\simeq v_{\parallel} \simeq c$, and
  \be
  \label{velocity1}
    v_{\perp} \simeq c\cdot \left(\frac{\rm 10^6~cm}{R_{\rm NS}}\right)^{1/2} \cdot \left(\frac{M_{\rm NS}}{2~M_{\odot}}\right)^{1/2}.
     \ee
  In what follows it is more convenient to parametrize the velocity of an  AQN when it enters the NS atmosphere  in terms of the proper $\eta_\mu$ velocity and 4-momentum $p_\mu$ defined  in the usual way:
\be
 \label{proper}
\eta_\mu=\gamma(c,\vec{v}), ~~~~~~~ \gamma=1/{\sqrt{1-v^2/c^2}}, ~~~~~~~~ p_{\mu}={\cal M}_0\cdot \eta_{\mu},
 \ee
  where ${\cal M}_0\simeq m_pB$ is the AQN's rest mass expressed in terms of the proton mass as reviewed in section \ref{sec:QNDM}. The key observation here is that  the Mach number $M\gg 1$ is always   large   for a typical AQNs entering the NS atmosphere,
\beq
 \label{Mach1}
 M &\equiv  &\frac{\sqrt{v^2_{\perp}+ v^2_{\parallel}}}{c_s} \\
&\simeq &  20\cdot  \left({ \frac{1~ {\rm MeV}}{{\rm k_B}T}}\right)^{1/2}\cdot  \left(\frac{\rm 10^6~cm}{R_{\rm NS}}\right)^{1/2} \cdot \left(\frac{M_{\rm NS}}{2~M_{\odot}}\right)^{1/2} \nonumber 
 \eeq
is much larger than one. One can also consider the Alfv\'{e}n Mach number $M_A$ defined in conventional way as:
 \be
 \label{Mach-alfven}
 M_A \equiv \frac{\sqrt{v^2_{\perp}+ v^2_{\parallel}}}{v_A},
 \ee
 where the Alfv\'{e}n velocity $v_A$ is given by eq.(\ref{definitions}). As we discussed in section  \ref{SP} the Alfv\'{e}n velocity $v_A\simeq  c$
  is very close to the speed of light  in NS environment.  At the same time 
  the nugget's velocity $v_{AQN}\sim c$  is only a fraction of the speed of light.   Therefore $M_A< 1 $ in general.
  However, in close vicinity of the reconnection region where the magnetic field strongly fluctuates the Alfv\'{e}n Mach number $M_A>1$ could be larger than one (locally), depending on specific  details, including the initial direction velocity of the nugget. Indeed, the magnetic field must change the direction  at the point of reconnection, and therefore it must vanish (locally) at some point. Therefore, the global characteristic (\ref{beta1}) does not reflect the local features of the system where $\beta$ could be many orders of magnitude larger due to the strong (local) fluctuations of the magnetic field in the vicinity of    (would be) the reconnection area.

  In a former  case when $M \gg 1$ and $M_A<1$ one should expect the so-called ``slow shock wave", while  in a latter case when $M \gg 1$ and $M_A>1$, very close to the reconnection region, 
  one should expect the ``fast shock waves" \cite{Landau}.
  
%  where the Alfv\'{e}n velocity $v_A$ is given by eq.(\ref{definitions}). As we discussed in section  \ref{SP} the Alfv\'{e}n velocity $v_A\sim  c$, similar %to the nugget's velocity $v_{AQN}\sim c$ as they both are
%   close to the speed of light in NS environment. Therefore $M_A\sim 1 $, and it  could be smaller or larger than unity, depending on specific  details, %including the initial direction velocity of the nugget. We assume in what follows (similar to   our arguments in section \ref{m_reconnection}) that a %finite portion of the nuggets may have $M_A>1$, in which case one should expect very strong shock as $M \gg 1$ in spite of the fact that parameter %$\beta$ is typically, on average, very small in NS environment. 
   
%   The key element in our arguments is that the region  where reconnection may start has a very complicated structure with very large changes  of the %magnetic field. In fact, the magnetic field must change the direction  at the point of reconnection, and therefore it must vanish (locally) at some point. %Therefore, the global characteristic (\ref{beta1}) does not reflect the local features of the system where $\beta$ could be many orders of magnitude %larger due to the strong (local) fluctuations of the magnetic field in the vicinity of    (would be) the reconnection area. 

\subsection{FRB'S relevant parameters from the AQN-framework   perspective}\label{FRB_parameters}

  In this subsection we shall argue that
  the key parameters (which  have been estimated  in \cite{kumar_1, kumar_2}  by assuming that FRBs   are powered by magnetic reconnections in NSs) are naturally emerging from our framework. The  procedure used in    \cite{kumar_1, kumar_2}
  can be treated as the bottom-top scenario when all  key parameters are   extracted from the observations rather than from a theoretical computation based on the first principles (in which case it would be considered  a top-bottom scenario).

In our estimates above, we used two important parameters (magnetic field ${\cal{B}}$ and electron density $n$), hinting to specific features of the  environment for FRB explosions to occur. In this subsection we consider few additional  key  elements  which were extracted  from \cite{kumar_1, kumar_2} by assuming   that the FRBs are powered by the antenna mechanism when the observed FRB's   radiation in radio wave bands is generated by the so-called curvature radiation.

  Before we proceed with the estimates we have to introduce several relevant elements along the line advocated in \cite{kumar_1, kumar_2}. To produce the observed FRB luminosity $L_{\rm iso}\sim 10^{43} {\rm erg ~s^{-1}}$, electrons must form bunches and radiate coherently. The size of a bunch in the direction of the line of sight must not exceed
  $\lambda/(2\pi)\simeq 4.8 \nu_9^{-1} {\rm cm}$. The radiation formation length is $\rho/\gamma$, where $\rho$ is the local curvature radius of the $B$ field line and $\gamma$ is the Lorentz factor, which are defined in the conventional way. The frequency of the radiation
  and the single particle curvature power  are given by the conventional formulae \cite{Jackson}:
  \be
  \label{Jackson_definitions}
  \nu=\frac{c}{2\pi\rho}\gamma^3, ~~~  P_{\rm curv}\simeq \frac{2 e^2 c}{3\rho^2} \gamma^4,
    \ee
  When the particle is moving towards the observer, the isotropic equivalent luminosity is given by \cite{kumar_1, kumar_2}:
  \beq
  \label{luminosity}
  L^{\rm bunch}_{\rm iso} &\simeq  &\gamma^4   N_{\rm coh}^2   P_{\rm curv} \\
& \simeq & (1.8\cdot 10^{40} {\rm erg ~s^{-1}})\eta^2 n_{18}^2\nu_9^{-2}\rho_5^2  \nonumber \\
  {\rm where } ~~ \gamma &\simeq  &28 \nu_9^{1/3}\rho_5^{1/3}. \nonumber
  \eeq
 The maximum number of particles in one coherent bunch is determined by the condition  \cite{kumar_1, kumar_2}:
   \be
   \label{bunch}
   N_{\rm coh}\simeq \pi n \gamma^2 \left(\frac{\lambda}{2\pi}\right)^3\simeq 2.8 \cdot 10^{23} n_{18}\nu_9^{-7/3}\rho_5^{2/3}.
   \ee
   In estimates (\ref{luminosity}) and (\ref{bunch}) we used the same number density $n=10^{18}n_{18}~ {\rm cm^{-3}}$ which enters
   our formula (\ref{beta1}) for estimates of the parameter $\beta$, and expressed all relations in terms of the observable frequency
   $\nu_9$ defined as $\nu=\nu_9 ~{\rm GHz}$ using relations (\ref{Jackson_definitions}). The factor $\eta>1$ in eq.  (\ref{luminosity})
   accounts for non-coherence of the particles in the transverse direction when the intensities (rather than the amplitudes) of the radiation from different bunches must be added.

   The key dimensional parameter in these estimates is  the  local curvature radius $\rho$   which should be of order $\rho\sim 10^5~ {\rm cm}\sim 1~ {\rm km}$  to reproduce the observed FRB's luminosity  (\ref{luminosity}). This parameter has been extracted  from bottom-top analysis in \cite{kumar_1, kumar_2}   to match the observed features of the FRB radiation. From the perspective of the AQN framework the parameter $\rho $  should be treated as a typical distance where  the nuggets are capable of {\it coherently}  generating the shock wave on a time scale $ \Delta t$, which must be shorter than the observed $1~ {\rm ms}$ duration of FRBs. Precisely this shock wave initiates the magnetic reconnection
   on huge distances of order $\rho $  which eventually powers FRBs.
   
   Having estimated the relevant dimensional parameters such as $n=10^{18}n_{18}~ {\rm cm^{-3}}$ and $\rho\sim 10^5~ {\rm cm}\sim 1~ {\rm km}$
   one may wonder if these parameters are consistent with AQN framework advocated in the present work. To be more specific, the question we want
   to address is as follows. The proton column density  corresponding to these parameters is $\rho n\sim 10^{23} {\rm cm^{-2}}$. This is many orders of magnitude larger than the proton column density  for the solar corona $\sim 10^{18} {\rm cm^{-2}}$ where most of the AQNs get completely annihilated as reviewed in Sect. \ref{AQN-flares}. Is there any inconsistencies there?
    
   One should emphasize that there is no contradiction between these parameters in two different systems because the magnitude of the column density does not  completely describe the systems. The effective cross section is also a crucial element in the estimates of the annihilation rate. In fact,  the interaction of the AQNs in these  two very different environments  is also drastically different. Indeed,  the large effective cross section in the solar system is due to the long range Coulomb  forces for non-relativistic AQNs with typical velocities $v_{\rm AQN}\sim 10^{-3}c$. It should be contrasted with AQN velocities  in NS environment with $v_{\rm AQN}\sim c$ as estimated in \ref{sect:Mach1}. As the Coulomb cross section behaves as $1/v^4$ the effective interaction  with solar protons  from plasma is  much more efficient than with protons from  NS' s atmosphere.  Essentially, this implies that the protons from corona  can be effectively captured  by AQNs, which greatly increases the annihilation rate in the solar corona.
   
   A relatively mild increase of the  AQN  ionization charge $\sim T^{3/2}$ due to the temperature increase from $T\sim 10^2$ eV in the solar corona to $T\sim 10^6$ eV in NS does not modify the main conclusion that nuggets can easily enter the relatively dense region with $n\sim 10^{18}{\rm cm^{-3}}$ developing the shock waves.  A complete annihilation is likely to occur much later when the nuggets enter the    very dense  regions close  the NS crust. This  qualitatively differs  from the solar system where the AQNs are expected to get annihilated in the transition region with the altitude around 2000 km above the photosphere.
   
Now we return to our analysis of  the  time scale $ \Delta t$  where the nuggets are capable to {\it coherently}  generate  the shock waves. The  $ \Delta t$ can be estimated as follows:   

   \beq
   \label{time}
   \Delta t &\simeq & \frac{\rho}{v_{\perp}}\simeq 0.3\cdot 10^{-5}~{\rm s}\cdot \left(\frac{\rho}{10^5~{\rm cm}}\right) \cdot \left(\frac{R_{\rm NS}}{\rm 2.9~km}\right)^{1/2} \nonumber \\
   & \times & \left(\frac{M_{\odot}}{M_{\rm NS}}\right)^{1/2}  \ll   \tau_{\rm FRB},
   \eeq
 where $v_{\perp}$ is determined by (\ref{velocity1}) and $\tau_{\rm FRB}\simeq 10^{-3} {\rm s}$ is the FRB duration. After the time $\Delta t $ the AQN enters the very dense regions  of the NS and ceases  to produce any radiation signature visible from outside the NS.

The time scale $\Delta t$ in estimate (\ref{time}) should not be confused with
 a typical duration of the FRBs. Rather, this time scale (\ref{time}) plays the same role as the duration of a pre-flare stage in our analysis of the flares in solar corona. The development of the flare itself represents the next stage of the system's evolution. Similarly, the magnetic reconnection in FRBs is capable of developing over distances $\sim ~L$, which can be  estimated as follows
 \be
 \label{distance}
 L\simeq v_{\parallel}\Delta t\simeq  1~ {\rm km}\cdot \left(\frac{b}{b_{\rm cap}}\right) \cdot \left(\frac{\rho}{10^5~{\rm cm}}\right),
 \ee
 where $v_{\parallel}$ is determined by (\ref{velocity}). This distance $L$ should be interpreted as a typical distance parallel to  the NS surface where the magnetic field's configuration is prepared for the reconnection and the AQN can initiate   this reconnection
  by playing the role of a trigger along the  trajectory $L$. This scale plays the same role as $L$ in SP analysis (see section \ref{SP}).

  The total area $A$ which is potentially the subject of the magnetic reconnection is determined by the propagation of the shock wave front and the speed of sound, and can be estimated as follows:
  \beq
  \label{area}
  A&\simeq &L\cdot (c_s \Delta t) \\
&\simeq &  0.1~ {\rm km^2}   \left(\frac{{\rm k_B} T}{1~ {\rm MeV}}\right)^{1/2}  \left(\frac{b}{b_{\rm cap}}\right)  \left(\frac{\rho}{10^5~{\rm cm}}\right)^2 \times \nonumber \\
&&  \left(\frac{R_{\rm NS}}{\rm 2.9~km}\right)^{1/2}
    \left(\frac{M_{\odot}}{M_{\rm NS}}\right)^{1/2}, ~~~ \nonumber
  \eeq
  where we use expression (\ref{sound1}) for the speed of sound $c_s$. This area represents a small but finite portion $\sim \epsilon$ of the NS surface,
  estimated as
  \be
  \label{epsilon}
  \epsilon\simeq \frac{A}{4\pi R^2_{\rm NS}}\simeq 10^{-4} ~~{\rm for} ~~R_{\rm NS}\simeq 10~ {\rm km}.
  \ee
 The total magnetic energy $E^{\rm tot}_{\rm mag}(A)$ available for a FRB due to the  magnetic reconnection from the surface area $A$ can be estimated as follows
 \beq
 \label{m_energy}
 E^{\rm tot}_{\rm mag}(A)&\simeq&\frac{\epsilon}{8\pi}  \int d^3x {\cal{B}}^2 \nonumber \\
&\simeq & 10^{40}{\rm erg} \left(\frac{{\cal{B}}_{\rm surf}}{10^{14}~{\rm G}}\right)^2\left(\frac{R_{\rm NS}}{10~ {\rm km}}\right)^3 \\
 {\rm where}~~ {\cal{B}} &\simeq  &{\cal{B}}_{\rm surf}\frac{R_{\rm NS}^3}{r^3}.~~~~\nonumber 
 \eeq
 This estimate unambiguously shows that the magnetic energy (\ref{m_energy}) from a small area (\ref{epsilon})  is more than sufficient to power FRBs with
 typical luminosity (\ref{luminosity}), generating a total energy $  \sim10^{37} {\rm erg}$ over a duration time of order  $\sim 10^{-3} {\rm s}$.

One should comment here that the total energy of a FRB in our framework is proportional to the area $A$ which will be swept
 by the passing shock wave. Shock waves will always develop due to the large Mach number (\ref{Mach1}).
 They can initiate the FRBs when the AQNs hit an active area of strong  magnetic field gradient prepared for reconnection. The corresponding probability is suppressed by a small geometrical factor $A/L_{\perp}^2$, in close analogy with arguments presented in section \ref{scaling}  in relation with the  solar flare analysis. This simple geometrical argument suggests the   scaling $dN\sim E^{-\alpha}dE$ with
 $\alpha\simeq 2$ is consistent with Fig \ref{Shibata:2016} covering the  gigantic   energy interval for solar and star flares (see also footnote \ref{deviation} with  some comments).

One should expect that the exponent $\alpha\simeq 2$ for the FRB scaling  should be very similar to our analysis in section \ref{scaling} related to  the solar flares, as in both cases the scaling has a pure geometrical origin.   The observable data (\ref{FRB_scaling}) support this prediction of our framework when FRBs are powered by magnetic  reconnection triggered by AQNs. Furthermore, the exponent $\alpha\simeq 1.8$ quoted  in (\ref{FRB_scaling}) is consistent with $\alpha\simeq 1.8$  presented on  Fig \ref{Shibata:2016}, covering enormous range of energies for both solar and star flares. It is very likely  that the nature of the deviation from a simplified argument given above (suggesting $\alpha\simeq 2$) in both cases, is related to the same physics, and we shall not elaborate on the possible source for this deviation in the present work. It is expected that the amplitude of the frequency of appearance of the solar flares   presented on  Fig \ref{Shibata:2016} and FRB's  quoted  in (\ref{FRB_scaling}) should be different
 \footnote{In particular, the time scale to  ``prepare"  the magnetic field configuration
 for magnetic reconnection leading to the burst in NS or the Sun is drastically different. Furthermore,  the number of AQNs entering the NS and the Sun, which may trigger the explosions,  is  also very different.}. However, as we shall see in Section 5, an overall shift along the ``y'' axis of Fig \ref{Shibata:2016} for energies $E_{\rm iso}\in (10^{37}, 10^{40})~ {\rm erg}$ will match the FRB data quite well.
%In other words, the plot on   Fig \ref{Shibata:2016}
% cannot be simply extrapolated to larger energies accounting for FRBs. Instead, it can be extended to higher  energies corresponding to FRBs with a shift along the ``y" axis.  More specifically, the corresponding entire part of the plot with energies $E_{\rm iso}\in (10^{37}, 10^{40})~ {\rm erg}$ from (\ref{FRB_scaling}) must be shifted down similar to the superflares portion shown on Fig \ref{Shibata:2016}.

  Another important parameter, which was estimated in the bottom-up analysis of \cite{kumar_1, kumar_2}, is the required inflow speed of reconnection, $\beta_{\rm in}c$, required to maintain the coherent emission of the bunch and match the observed FRB luminosity. The corresponding estimate of $\beta_{\rm in}$  gives a specific and unique prediction of the FRB duration $\tau_{\rm FRB}$ in the AQN framework as follows:
  \beq
  \label{duration}
 \beta_{\rm in} &\leq &\frac{1}{4\gamma} ~~~ \Rightarrow ~~~~ \tau_{\rm FRB}\approx \frac{L}{\beta_{\rm in}c}\geq \frac{4\gamma L}{c} \\
&\simeq & 0.4\cdot 10^{-3} {\rm s}\cdot \left(\frac{b}{b_{\rm cap}}\right) \rho_5^{4/3}\nu_9^{1/3}, \nonumber 
  \eeq
  where $L$ is determined by eq. (\ref{distance}) and interpreted as a typical distance where the magnetic field's configuration will be reconnected due to the AQN passing through    this region
   and initiating the reconnection along the  trajectory $L$. In the estimate (\ref{duration}) we express $\gamma$ in terms of the frequency $\nu$ and curvature radius $\rho$ according to (\ref{luminosity}). The estimate (\ref{duration}) is a highly nontrivial consistency check for our proposal as it includes parameters from the AQN framework  as well as parameters  extracted in \cite{kumar_1, kumar_2}  from a bottom-top analysis of FRBs.

   One should comment here that the estimate on $\beta_{\rm in}$ is based on an approximate formula for the induced electric field $E_{\parallel}$ parallel to the original static magnetic ${\cal{B}} $ field, i.e.
   \be
   \label{E}
   E_{\parallel}\simeq {\cal{B}}\sin\theta_{\cal{B}}\beta_{\rm in}, ~~~~~ \vec{E}\cdot \vec{{\cal{B}}}\sim {\cal{B}}^2\sin\theta_{\cal{B}}\beta_{\rm in}.
   \ee
   The generation of $E_{\parallel}$ is a required feature of the magnetic reconnection  mechanism. Concurrently, this  electromagnetic configuration where $ \vec{E}\cdot \vec{{\cal{B}}}$ does not vanish according to (\ref{E}) describes the dissipation of the magnetic helicity $\cal{H}$ of the system as discussed in Appendix \ref{sect:helicity}. The exact term with $ \vec{E}\cdot \vec{{\cal{B}}}\neq 0$ is responsible for the FRB radiation. The same term describes a slow  dissipation   pattern of the Magnetic helicity.
   A numerical suppression of the dissipation term  $\sim \vec{E}\cdot \vec{{\cal{B}}}$, as observed in numerical studies reviewed  in Appendix \ref{sect:helicity}, manifests itself
   in smallness  of the factor $ \beta_{\rm in}\ll 1 $, as stated in (\ref{duration}).

One should mention that the duration $  \tau_{\rm FRB}$ is getting  shorter when the typical length scale $L$ is getting smaller.
   It happens when the  impact parameter $b$ is becoming smaller. In the AQN framework, it corresponds to a smaller area $A$ where the reconnection occurs (\ref{area}), which unambiguously implies a smaller luminosity for FRB (which is directly proportional to the area according to (\ref{m_energy})). Therefore, in the AQN framework, a weaker  luminosity $L_{\rm iso}$ corresponds to a  shorter burst duration $\tau_{\rm FRB}$. A similar estimate   \cite{Zhitnitsky:2018mav} for a duration of a solar flare would give much longer time scale measured in hours rather than in milliseconds (\ref{duration}) because the Alfv\'{e}n velocity  $v_A\sim 10^{-3}c$ is much smaller in the solar corona than in a NS where $v_A\sim c$. A typical length scale $L$ in the corona is measured in thousand of kilometres
   rather than $1~ {\rm km}$ scale, entering the estimate (\ref{duration}). This six orders of magnitude difference is translated  to drastic changes in time scales from milliseconds bursts in NS to hours for solar flares. Nevertheless, the physics in both cases is very similar, and the scaling relation (\ref{FRB_scaling}) with $\alpha\simeq 1.8$ as discussed above supports this conjectured similarity.

Finally, the AQN framework gives a very reasonable ratio for the energy release in solar flares in comparison to FRBs.
  In both cases the energy release is proportional to the  strength of the magnetic field and the area which is swept by the passing nuggets igniting the reconnection. A very simplified  estimate can be expressed  as follows:
  \be
  \label{ratio}
  \frac{E_{\rm Solar ~ flare}}{E_{\rm FRB}}\sim \left(\frac{10^3 ~{\rm G}}{10^{14} {\rm G}}\right)^2\cdot\left(\frac{10^5~ {\rm km}}{1~{\rm km}}\right)^2\cdot\left(\frac{h_{\rm Sun}}{h_{\rm NS}}\right)\cdot \gamma^2\sim 10^{-6},
  \ee
  where $h_{\rm Sun}/h_{\rm NS}\sim 10^4 {\rm km}/10 {\rm km}= 10^3$ are the thickness for the magnetospheres for the Sun and NS correspondingly. In formula (\ref{ratio})    we also included the $\gamma\sim 30$ factor  to account for the beaming corrected total energy $ E_{\rm FRB}\sim E_{\rm iso}\gamma^{-2}$, instead of its isotropic equivalent. The estimate (\ref{ratio}) is quite reasonable as a typical solar flare is characterized by an energy $ \sim 10^{30} {\rm erg}$ to be compared with the typical  beaming corrected total energy of the FRB $\sim 10^{36} {\rm erg}$.

%-------------------------------------------------------------------
\section{AQN predictions  confronting  the FRBs observations}\label{morphology}

We have seen in the previous section that the AQN model provides a framework consistent with the theory developed by \cite{kumar_1,kumar_2}. In this Section, we are confronting our model to the currently known FRB statistics. This is an extremely difficult task because it is only the beginning of a new era of systematic observation of FRBs (e.g. see Figure 2 in \cite{2018NatAs...2..865K}). The statistics of current data is small and surveys vary greatly in sensitivity, field-of-view and frequency coverage.
At this time, there are six telescopes that have discovered FRBs, one which, CHIME \footnote{https://chime-experiment.ca/} has only one single discovery and another, UTMOST \footnote{https://astronomy.swin.edu.au/research/utmost/} ,only six. We therefore focus on the other four, which have better statistics, and, for three of them, relatively similar frequency coverage. Table \ref{table:data} lists the four telescopes.

Our model can make predictions on the energy distribution, burst duration and occurrence rate. As discussed in the previous Section, the magnetic field sourcing FRBs must be strong enough, larger than $10^{14}$ G, so that only magnetars can be progenitors. We assume that FRBs have a cosmological origin. If our model can explain single bursts and the repeating FRB121102, it should make predictions that are verified independently by all FRBs. All single bursts observed so far, and the Arecibo data for FRB121102, can be found in a compiled catalogue \footnote{http://www.frbcat.org}. The GBT-BL data for FRB121102 can be found in \cite{2018ApJ...863....2G}. 

\begin{table}
	\centering
	\caption{\label{table:data}Summary of the four data sets used for our comparison. ASKAP and GBT-BL correspond to the Australian Square Kilometer Array Pathfinder and Green Bank Telescope-Breathrough Listen respectively. The beam shape is given in arcmin$^2$, $\nu_c$ is the central frequency and $\Delta_\nu$ the bandwidth. References correspond to papers where the data used in this work can be found. The last two entries, Arecibo and GBT-BL, are data of the repeater FRB121102. The first two entries are the single bursts surveys.}
	\begin{tabular}{lccr} % four columns, alignment for each
		\hline
		Telescope & Beam & $\nu_c$  & $\Delta_\nu$ \\
         & [arcmin$^2$] &   [MHz] &   [MHz]\\
		\hline
		Parkes \cite{Petroff-2016,2018MNRAS.475.1427B} & 7.5$\times$7.5 & 1352 & 338\\
		ASKAP \cite{2018arXiv181004353M} & 30$\times$6 & 1297 & 336\\
        Arecibo \cite{2016Natur.531..202S} & 1.75$\times$1.75 & 1375 & 322\\
        GBT-BL \cite{2018ApJ...863....2G} & 2.2$\times$2.2 & 6400 & 4000\\
		\hline
	\end{tabular}
\end{table}

- {\bf Energy distribution}: 
An estimation of the FRB energy distribution can be obtained from the observed peak flux density, duration of the burst and estimated redshift. We should emphasize that the derived quantities are subject to uncertainties. For instance, in the case of single radio bursts, the position of the source is unknown within the beam. As a consequence, the measured fluence $F_{\rm obs}$ should be treated as a lower bound of the true value. We summarize below the steps that lead to an estimated FRB energy $E_{\rm FRB}$:

\begin{itemize}
\item The observed fluence $F_{\rm obs}=S_{\rm peak} \times W_{\rm obs}$, where $S_{\rm peak}$ and $W_{\rm obs}$ are the observed peak flux density, in [Jy], and width, in [ms], of the FRB respectively.

\item A redshift of each FRB is estimated, from the excess dispersion measure ${\rm DM}_{\rm excess}={\rm DM}_{\rm FRB}-{\rm DM}_{\rm gal}$, where ${\rm DM}_{\rm FRB}$ is the measured ${\rm DM}$ and ${\rm DM}_{\rm gal}$ is the estimated ${\rm DM}$. The redshift is given by $z={\rm DM}_{\rm excess}/1200 ~ {\rm pc ~ cm}^{-3}$, where the denominator is the estimated inter-galactic medium electron density \cite{ioka-2003}. This estimate is an upper bound of the real redshift because it neglects the contribution of the FRB host galaxy to the measured DM.

\item The energy of the FRB, $E_{\rm FRB}$ is estimated from the fluence $F_{\rm obs}$, the observing bandwidth $\Delta_\nu$ and the luminosity distance $D_L(z)$:

\begin{equation}
\label{energy}
E_{\rm FRB}=F_{\rm obs} \times \Delta_\nu \times D_L(z)^2 (1+z) 10^{-26} ~ {\rm erg}
\end{equation}
\end{itemize}

\begin{figure}
\centering
\includegraphics[width=0.4\textwidth]{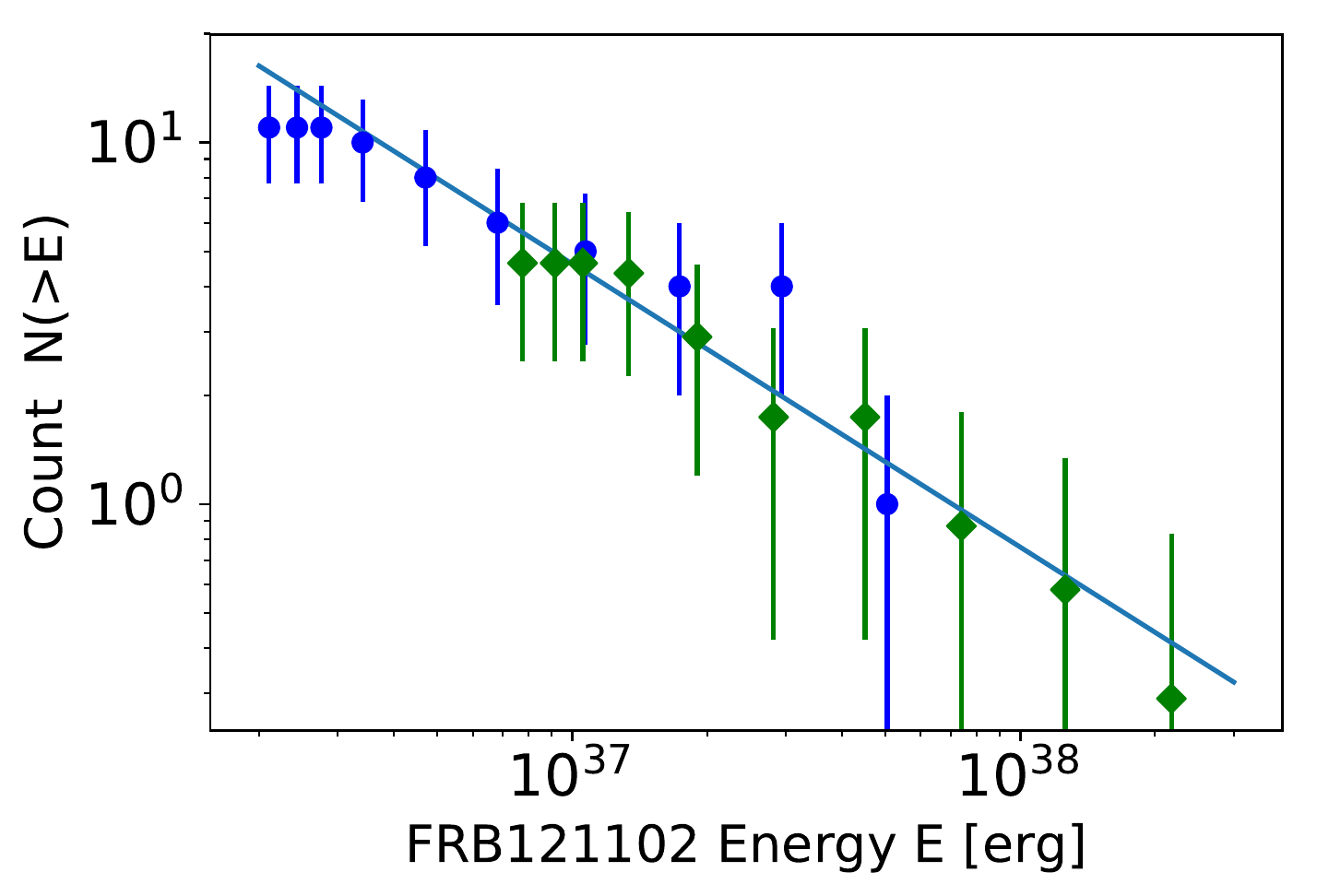}
\includegraphics[width=0.4\textwidth]{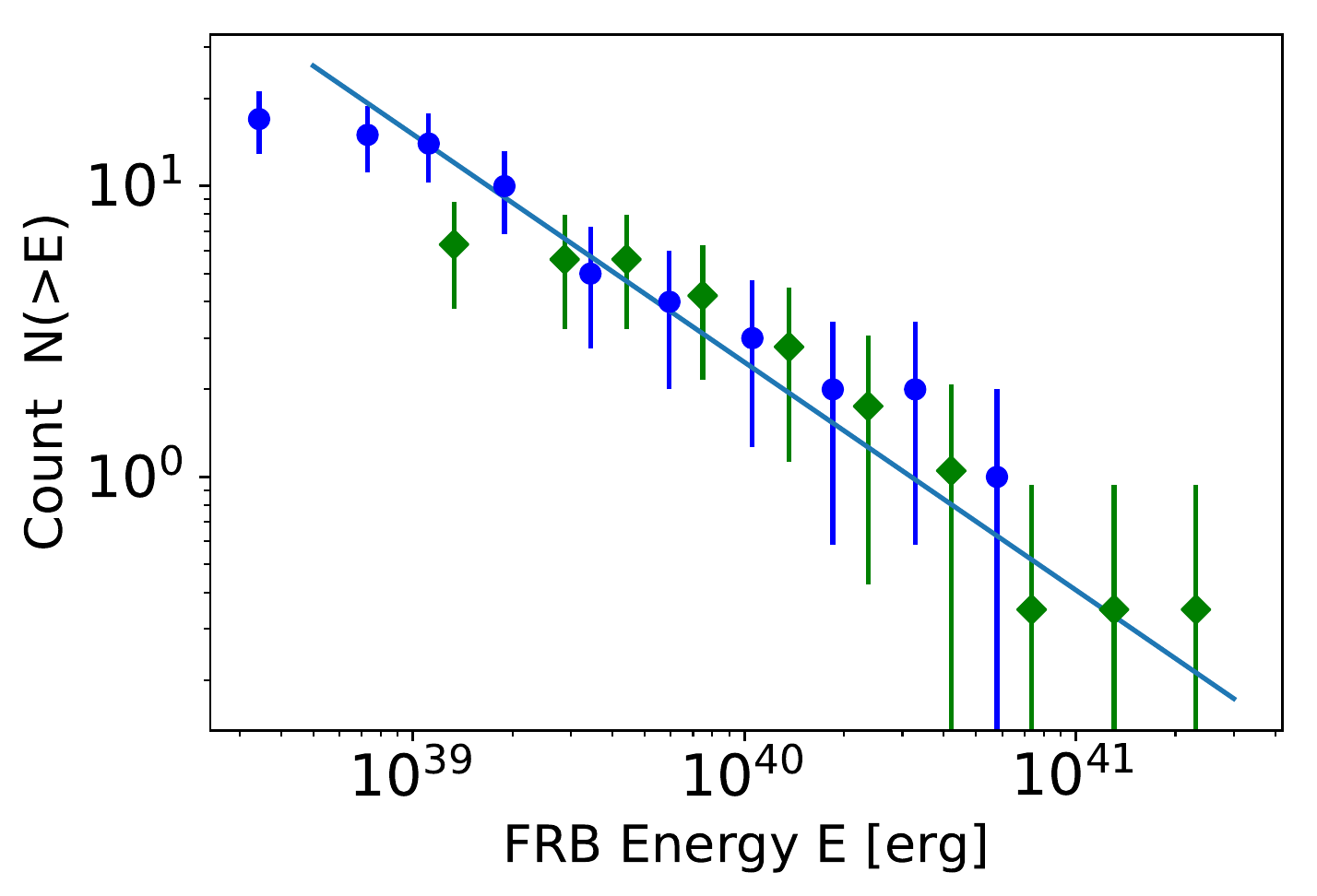}
\caption{\label{counts:stats} Top panel: FRB cumulative count for the energy of the repeating FRB121102, from Arecibo  (blue filled circles) and the GBT-BL (green filled diamonds). The solid line shows the best fit to the Arecibo points. The amplitude of the GBT-BL points has been adjusted to align with the solid line, but the slope was not changed. Bottom panel: FRB cumulative count of the derived energy $E_{\rm FRB}$ for single bursts. The blue circles and green diamonds correspond to Parkes and ASKAP data respectively. In order to limit the uncertainty from the Galaxy dispersion measure, only FRBs with galactic latitude $b> 20 ~{\rm deg}$ have been used here. The solid line is constructed from the best fit of the top panel, the slope is unchanged, only the amplitude is adjusted to fit the distributions.}
\end{figure}

Where $F_{\rm obs}$ is in Jy$\cdot$ms, $\Delta_\nu$ in Hz and $D_L(z)$ in cm.
We are assuming a power law model $dN(E) \propto E^{-\alpha}dE$, which cumulative counts are given by $N(>E) \propto E^{1-\alpha}$. For all FRBs, energy is calculated from the fluence using Eq. \ref{energy}. The top panel of Fig. \ref{counts:stats} shows the cumulative energy count $N>E_{\rm FRB}$ for the repeating FRB121102. The solid line shows a power law fit to the filled circle points which correspond to the Arecibo data \cite{2016Natur.531..202S}. The slope of the fitted line is $\alpha=1.78\pm 0.07$. The filled diamond points on the same panel show the GBT-BL data \cite{2018ApJ...863....2G} superimposed to the plot. A rescaling factor of $0.3$ was applied to the amplitude to align the data with the Arecibo fit. Although the central frequency is five times higher for GBT-BL than for Arecibo (see Table \ref{table:data}), and the badwidth $\Delta_\nu$ is ten times larger, the two distributions have a similar power law slope. Even more surprisingly, the same slope seem consistent with single bursts events. The bottom panel in Fig. \ref{counts:stats} shows the cumulative counts for single bursts FRBs. The filled circle and filled diamond points correspond to the Parkes \cite{Petroff-2016,2018MNRAS.475.1427B} and ASKAP \cite{2018arXiv181004353M} data respectively. The solid line is not a new fit, it is the fitted power law from the top panel, with amplitude scaled to align the data points with the line. The x-axis of the bottom panel is the derived FRB energy $E_{\rm FRB}$. This is a highly indirect quantity which depends on the dispersion measure and fluence as shown in Eq. \ref{energy}. The fact that the same power law seem to describe both FRB121102 and the single bursts is remarkable.  

It is also remarkable that the slope is consistent with the Solar Flare energy distribution shown on Figure \ref{Shibata:2016}. This coincidence has a natural explanation in our model: the Solar flares and FRBs are phenomena caused by the magnetic reconnection ignited by the incoming AQNs and, according to \ref{scaling}, the energy distribution slope $\alpha$ has a simple geometrical origin which holds for all systems, including solar flares, solar  microflares and FRBs.  This generic prediction of the  model is the direct consequnce of our  framework when AQNs play the role of the triggers which ignite the large flares in the Sun or FRBs in the NS.

\begin{figure*}
\centering
\includegraphics[width=1.\textwidth]{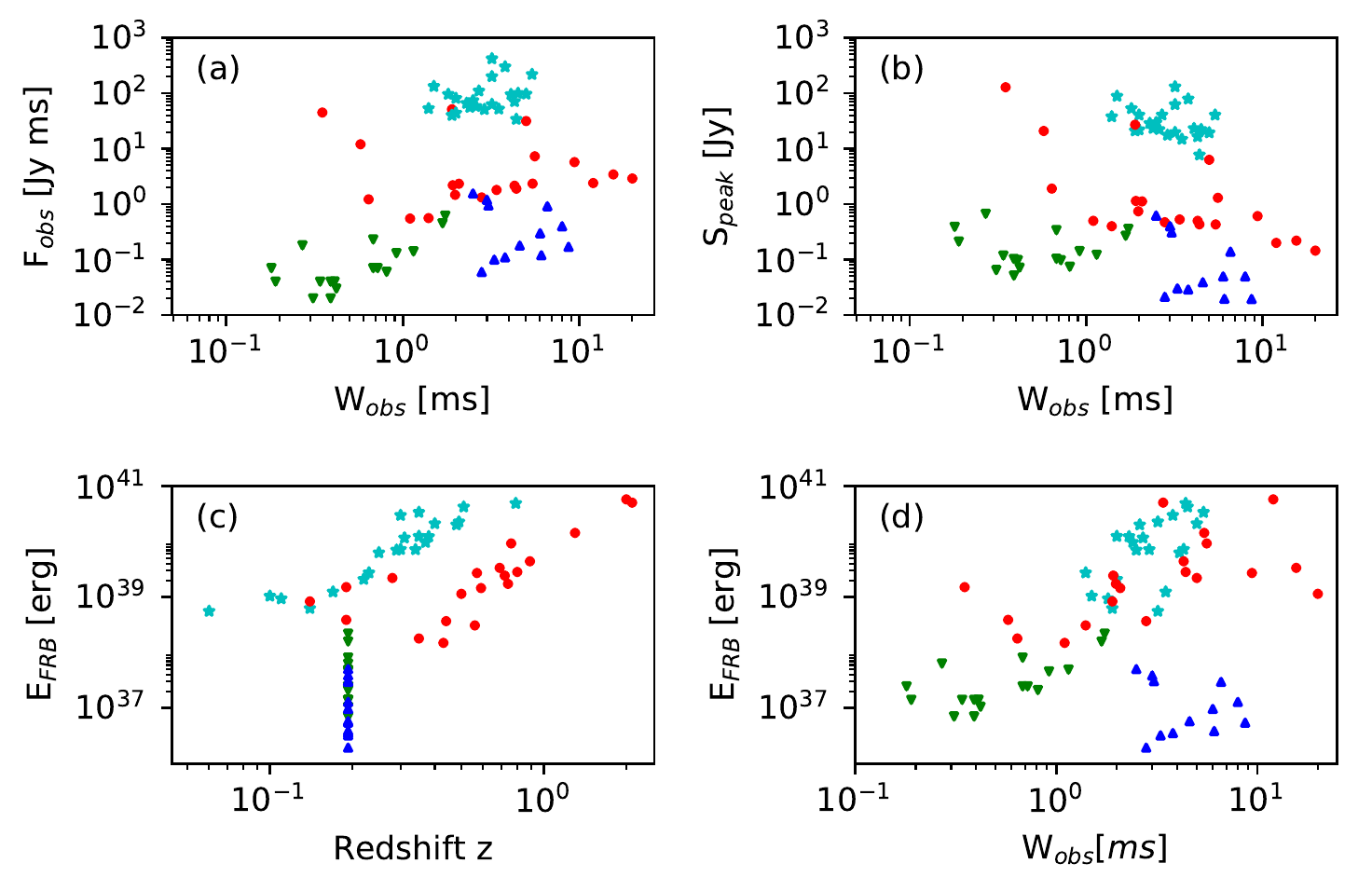}
\caption{\label{datacombo} Statistics of the four surveys used in this paper and listed in Table \ref{table:data}. In all panels, cyan stars, red circles, blue up-triangles and green down-triangles are for the ASKAP, Parker, Arecibo and GBT-BT surveys respectively. ASKAP and Parkes apply to single bursts, and Arecibo and GBT-BL apply to the repeating FRB121102. Panel (a) shows the Fluence $F_{\rm obs}$ versus burst duration $W_{\rm obs}$. Panel (b) shows the burst peak intensity $S_{\rm peak}$ versus $W_{\rm obs}$. Panel (c) shows the estimated burst energy $E_{\rm FRB}$ versus the estimated redshift $z$ using Eq. \ref{energy}. The redshift of FRB121102 is z=0.193, which is the redshift of the dwarf galaxy associated with FRB121102. Panel (d) shows $E_{\rm FRB}$ versus $W_{\rm obs}$.}
\end{figure*}

%the intensity distribution and the statistical properties of FRBs was calculated in many papers (e.g. \cite{Li-2017} or \cite{Katz-2016} for a review). If our proposal is correct, that AQNs are the triggers of flares in magnetar's atmospheres. We   expect that the FRB  energy distribution should be  similar to the one observed for the solar flares  shown on Fig. \ref{Shibata:2016}, extrapolated to much higher energy. In order to visualize the low statistics of FRBs, we will calculate the cumulative counts $N(>E)$ instead of the FRB frequency $N(E)$. If we assume a power law model $dN(E) \propto E^{-\alpha}dE$, then $N(>E) \propto E^{1-\alpha}$, which is very similar to the exponent shown on Fig. \ref{Shibata:2016} for very different systems. It definitely supports our proposal that the  solar flares, the  superflares, and FRBs have a common origin when the exponent  $\alpha \sim 2$ is  the direct consequence of the AQN framework when all these events are triggered and sparked by the same DM nuggets with the same features.  

%The bottom panel in Fig. \ref{counts:stats} shows the cumulative counts for single bursts FRBs. The line on this panel is not a fit, it is the count slope measured from the top panel and overlaid on the bottom panel where only the amplitude is renormalized. It is remarkable that the slope is in agreement with the fluence distribution of FRB121102, suggesting a common physical origin between FRB121102 and single bursts. It is also remarkable continuity with the solar flare statistics shown on Fig. \ref{Shibata:2016}. 

It is definitely consistent with our proposal that the  solar flares, microflares, the  superflares, and FRBs have a common origin where the exponent  $\alpha \sim 2$ is  the direct consequence of the AQN framework when all these events are triggered and sparked by the same DM nuggets with the same features. It is a nontrivial prediction of our framework that all newly founded FRBs, even with different instruments, should follow the same energy distribution function. 

Note that, for the single bursts FRBs, we only kept the FRBs at galactic latitude $b$ higher than $|b| > 20~{\rm deg}$, in order to minimize the galactic contamination on the dispersion measure, but the result from the bottom panel of Fig. \ref{counts:stats} are only marginally changed if the lower galactic latitude FRBs are included. We keep in mind that the FRB energy count is still subject to large uncertainties. Beyond the uncertainty on $F_{\rm obs}$, the redshift estimate relies on the proper removal of the Galaxy and the host contribution to the dispersion measure, the latter is neglected in the current estimates. Consequently, luminosity distance and $E_{\rm FRB}$ are affected by not very well known errors. It would be particularly interesting to see if future data for single burst will continue to align with the repeating burst energy counts as shown in the lower panel of Figure \ref{counts:stats}.

- {\bf Flare intensity and duration}: Our estimate for the duration of the FRBs given by Eq.(\ref{duration}) is consistent with a typical FRB event. Furthermore,  according to section 4, the flare intensity should be proportional to its duration. The FRB opening angle is given by $2/\gamma$, and with $\gamma \sim 30$, this corresponds to a beam size of $\sim 0.066$ rad. The typical magnetar rotation period is approximately $1-10$ sec. It means that the time it takes for the FRB beam to sweep an observer on Earth is between 10 and 100 ms, but these number are upper bounds because they correspond to the optimal situation where the emission is pointing straight to the observer at the maximum intensity.
Given that the observed FRB duration is $\sim 1-10$ ms, which is comparable to the visibility window of the emission, it is expected that a fraction of the flaring events will be truncated, so not all the energy will be measured. This will happen when the emission starts before, or end after, the beam becomes visible. It is therefore not possible to check the relationship between flare intensity and duration without a much better statistics and a more sophisticated modeling of the geometrical configuration.

%Our contribution into this field can be summarized as follow: 
-{\bf FRB occurrence rate}: In our model, the FRBs correspond to flares in magnetar atmospheres. With a well defined progenitor population and the energy scaling given by Eq. (\ref{ratio}), it is possible to estimate an order of magnitude for the observed FRB occurrence rate. The argument goes as follows:

%\begin{itemize}
$\bullet$ \underline{Number of active magnetars:} Only one out of ten supernovae turn into an {\it active} magnetar. We follow \cite{2007MNRAS.381...52G} which estimate the magnetar birth rate to be $\sim 0.2$ per century, and a total number of active magnetars to be $N_{\rm mag}\simeq 20$ on average in a galaxy.

$\bullet$ \underline{Frequency appearance}: We need to know the observed number of flares a magnetar experience per unit of time. It is currently not possible to predict the intrinsic occurrence of flares because of the unknown physical conditions in the plasma, the effect of turbulence and the configuration of reconnection in the NS magnetosphere. 
We will infer an estimate based on our model, which relates FRBs to Solar flare energy via the scaling relation Eq. (\ref{ratio}). This equation is a prescription on how, in our model framework, to convert a magnetar flare energy in a Solar flare energies given the physical parameters at play (size of the progenitor, strength of the magnetic field and thickness of the atmosphere). We decided to use $10^{39}$ erg as the minimum FRB rest frame flare energy to estimate a rate. Flares below this cut will not be counted in our estimate. This value is justified by looking at the bottom panel in Figure \ref{counts:stats} which shows that the energy $\sim 10^{39}$ erg corresponds to a cut off where the single bursts surveys ASKAP and Parkes are loosing sensitivity. Corrected from relativistic beaming with $\gamma\simeq 30$ this corresponds to an energy of $\sim 10^{36}$ erg. Using Eq. (\ref{ratio}), this corresponds to a typical Solar flare energy of $\sim 10^{30}$ erg. Our argument is that the rate of falling AQNs triggering $\sim 10^{30}$ erg flares in the Sun is that rate that triggers $\sim 10^{36}$ erg flares in magnetars.
According to \cite{Hannah-2008}, there is, on average, $10-20$ solar flares per year above energy $E> 10^{30}$ erg; $10^{30}$ erg approximately identifies the formal separation between micro-flares and normal/super flares. This corresponds to an average value of $N_{\odot}(E>  10^{30}~{\rm erg})\simeq 0.04$ flare per day for the Sun, which, in our model, would be triggered by AQNs.

In order to apply this solar flare rate to the NS, we have to take into account the fact that there is far less AQN hitting a NS per unit of time than for the Sun, because the impact parameter for the former is smaller. The ratio of the number of AQNs   hitting  the Sun over   the number of AQNs  hitting the magnetar per second is determined by the ratio of impact capture parameters square, which
%The reason is that every second the magnetar will collect far less AQN particles than the Sun, and the reducing factor is proportional to the AQN swept up volume.
  is given by the square of Eq (\ref{capture1}), i.e. $\left( b_{\rm cap~NS}/b_{\rm cap ~\odot}\right)^2$. Putting the last two estimates together, we can write the daily rate $f_{\rm NS}$ of giant flare on a magnetar in a given galaxy as follows:

\beq
\label{galrate}
f_{\rm NS} &=& N_\odot(E>  10^{30}~{\rm erg})\cdot  \left({b_{\rm cap~ NS}\over b_{\rm cap ~\odot}}\right)^2 \cdot N_{\rm mag} \nonumber \\
&\simeq &8\times 10^{-6} ~ {\rm day}^{-1} \sim 0.003 ~ {\rm yr}^{-1}
\eeq

$\bullet$ \underline{Daily occurrence of FRBs:} The next step in our estimates is to compute the  FRB rate for the entire Universe. To accomplish  this  task we have to multiply $f_{\rm NS}$ by the number of observable galaxies $N_{\rm gal}\simeq 10^{11}$. The last step is to multiply by the probability the flare will be detected on Earth. We have seen before that, in our framework, the emission is strongly beamed and the $\gamma$-factor is of the order $\sim 30$, corresponding to beam width of $\delta\theta \sim 2/\gamma$. The chance that the beam points towards us is given by
\begin{equation}
{\Delta\Omega\over 4\pi} \simeq {\delta\theta^2\over 4\pi} \sim 3\times 10^{-4}
\end{equation}
The predicted daily occurrence $n_{\rm FRB}$ of FRBs can then be estimated as
\begin{equation}
\label{totalrate}
n_{\rm FRB}\sim N_{\rm gal}\cdot f_{\rm NS} \cdot {\Delta\Omega\over 4\pi}\sim 2.4\cdot 10^2~{\rm day}^{-1},
\end{equation}
The actual FRB occurrence rate is not yet well known, and it varies greatly as function of telescope sensitivities and selection effects \citep{2016MNRAS.455.2207R,2018Natur.562..386S}.
Moreover, the occurrence rate is usually quoted  for a given survey fluence limit and frequency band. Our approach is also different in that our estimated rate depends on an arbitrary FRB energy cutoff, and not an arbitrary Fluence cutoff. Our estimate (\ref{totalrate}) is nevertheless in broad agreement with current constraints on FRB rate \citep{2016MNRAS.455.2207R,2018Natur.562..386S}.

-{\bf FRB statistics}: Figure \ref{datacombo} shows FRBs statistics from the four surveys used in this study and summarized in Table \ref{table:data}. Panels (a) and (b) show how the measured quantities, burst duration $W_{\rm obs}$, peak intensity $S_{\rm peak}$, are related to the fluence $F_{\rm obs}$. One clearly see that the data for the repeating FRB121102 (up and down triangles) are significantly lower energy than all single bursts. It is also very clear from these two panels that the current surveys do not overlap in terms of sensitivity and that the selection function between them is very different. It is interesting to note that the estimated burst energy $E_{\rm FRB}$ is significantly less for FRB121102 than for single bursts. This is consistent with our framework where lower energy bursts are much more numerous than high energy ones, with power law $E^{-\alpha}$ and $\alpha\sim 1.78$. We therefore do not regard the lack repeating bursts, in association with single bursts, as in contradiction with our model. Panel (c) shows that the distribution of $E_{\rm FRB}$ for FRB121102 is indeed significantly below all single bursts energies. The GBT-BL data points (green down-triangles) correspond to much higher central frequency and larger bandwidths than the other measurements (see table \ref{table:data}. For this reason, it is difficult to compare GBT-BL to the other data sets. Nevertheless, it is interesting to notice that a reduction of $E_{\rm FRB}$ by a factor ten for the GBT-BL points (in order to adjust the GBT-BL bandwidth to a value close to ASKAP, Parkes and Arecibo) seems to align the GBT-BL points with Arecibo (blue up-triangles) data points, despite the difference of a factor five in their central frequency. This can be seen in panel (d) where lowering the down-triangles by a factor $\sim 10$ brings them in the continuity of the up-triangle points. Overall, Figure \ref{datacombo} shows a diversity in surveys sensitivities which still makes it difficult to infirm or confirm a common origin for all FRBs, but at this stage, none of the current observations are in contradiction with our framework.

%-------------------------------------------------------------------
\section{Conclusion and final remarks}\label{conclusion}
%\input{conclusion.tex}

%Magnetization and scattering located near the FRB source favors young stellar populations (magnetars, young supernovae remnant) (https://arxiv.org/abs/1512.00529)

We propose a scenario which provides a coherent emission mechanism for FRBs. It is based on the antenna curvature mechanism, developed form a bottom-up approach by \cite{kumar_1,kumar_2}, in the context of the Axion Quark Nugget model. In this scenario, FRBs are at cosmological distance and are emitted when an AQN from the surrounding dark matter environment falls into the atmosphere of a magnetar, triggering magnetic reconnection and initiating a giant flare. Our study shows that the energetics of the process is consistent with antenna curvature mechanism and with current FRB constraints.
%the antenna curvature emission; it provides an explanation for both the duration and the observed energy distribution. We predict a steep slope for the energy distribution, consistent with \cite{Law:2017} and with the CHIME pathfinder measurements \citep{Amiri1-2017}. 
Magnetars are the central objects in our scenario because very strong magnetic fields are required (${\cal B} > 10^{14}~G$) in order to maintain the emission mechanism coherent by confining electrons along ${\cal B}$ lines in the quantizing magnetic field.

Our contrubution into this field is very modest: we argued that AQNs 
play the role of the triggers which initiate large flare resulted from the magnetic reconnection, as previously argued in refs.\cite{kumar_1,kumar_2}. We identify these flares in NS    with FRBs. The important point is that this novel element of  our proposal unambigoulsy predicts the slope for the energy distribution as shown on Fig. \ref{counts:stats}. Furthermore, in our framework this slope must be the same for the solar microflares, large flares and superflares,  
which is consistent with   energy distribution shown on Figure \ref{Shibata:2016}.  

%Our scenario follows a proposal that AQNs can trigger and initiate flares in the Sun \cite{Zhitnitsky:2018mav}. The AQN dark matter model was invented as an explanation of the observed ratio $\Omega_{\rm dark} \sim \Omega_{\rm visible}$, and has no free fundamental parameter other than the Axion mass. \cite{Raza:2018gpb} showed that the AQNs moving through the coronal plasma (and annihilating) could both explain the EUV excess and the drastic changes of the temperature in the Transition Region, observed to be independent of the heliocentric location on the Sun. Remarkably, \cite{Raza:2018gpb} predict the correct missing energy input for the solar corona, and an energy injection altitude in agreement with the temperature and mass density profile of the solar atmosphere. \cite{Zhitnitsky:2017rop} showed that AQNs can also trigger larger solar flares. This mechanism is consistent with the observed scaling of the flare distribution $dN(E)$ as a function of the flare's  energy $E$.

FRBs is a rapidly changing field \cite{2018NatAs...2..865K}, but they remain mysterious objects. It is not yet possible to discriminate between the different FRB progenitors because of the limited statistics \cite{2018NatAs...2..842P}. Among the tests that could discriminate our AQN scenario from others, we would like to mention the following:

\begin{itemize}

\item The spectral signature will be a strong discriminator. So far, there is no known X-ray or high energy counterpart to FRBs. With the AQN model, it is in principle possible to predict a precise wavelength signature using MHD simulations, it is a major topic left for future studies. It is interesting to note that the antenna curvature mechanism in magnetar environment predicts a strong suppression of emission from scales smaller than the two-stream current instabilities \citep{kumar_2}. This corresponds to a cut off frequency in the mm/infrared wavelengths. The observation of direct counterparts at smaller frequencies would invalidate the antenna mechanism and therefore the AQN model as a source of FRBs.

\item Our model unambiguously predicts a correlation between the total energy flare and its duration. As discussed in Section \ref{morphology}, there are geometrical complications which have to be modeled in order to this prediction as a test, but this is a prediction unique to the AQN model.

\item Our AQN model does not make any distinction between single and repeating FRBs. Flare bursts should appear at random, when an infalling AQN hits a region with magnetic field prepared for reconnection. Our scenario therefore predicts that all FRBs should be repeating, the frequency should be driven by external conditions only (e.g. dark matter density, magnetic environment). So far, there is only one known repeating FRB, FRB121102. It is possible that a reason why other FRBs have not been seen repeating yet, might be because they were not observed long enough.
It should be noted that the highest energy of the FRB121102 bursts is $\sim 10^{39}$ erg, which is among the lowest energy of single bursts. If we assume that the repeating bursts of FRB121102 follow the same scaling $\sim E^{-1.78}$, lower energy flares should be much more frequent, and the rate Eq. (\ref{totalrate}) should be significantly higher. In the AQN model, single and repeating bursts have the same physical origin and their occurrence rate should match the same power low for both, as suggested in Fig \ref{counts:stats}. Better FRB statistics, e.g. with CHIME, will soon test this hypothesis.

\item We want to make few comments on polarization properties, which carries valuable information on the physical process at play. For instance, the recent measurement of high rotation measure in a single burst FRB \citep{Masui-2015} provides an independent indication that FRBs are at cosmological distance. Within our AQN scenario, the energy source comes from magnetic reconnection fed by the helical magnetic field as discussed after eq. (\ref{E}) and in Appendix \ref{sect:helicity}. The dissipation rate of the magnetic helicity $d{\cal{H}}/dt $ defined by eq. (\ref{time-derivative}) determines
 the duration $\tau_{\rm FRB}$ of the FRB  as given by eq. (\ref{duration}). It unambiguously implies that the FRB emission must be linearly polarized with polarization directed (locally) along the external static  magnetic field $\vec{{\cal{B}}}$ where FRB is originated. What then happens to the polarization is a difficult question to answer, because of the many cosmological uncertainties along the propagation path of the FRB emission before it reaches the Earth. The question deserves future studies and analysis before it can be used to test the AQN scenario.

\end{itemize}

%Finally, it is interesting to note that there are several similarities between Soft Gamma Repeaters (SGR) and FRBs regarding energetics and time scales (not for occurrence rate which is $\sim 10^4$ times higher for FRBs). Since SGRs are thought to be caused by giant flares, it has indeed been suggested that SGRs and FRBs have a common origin \citep{Katz-2016-2}. A recent study showed that the lack of temporal correlation between $\gamma$-rays and FRB emissions already rules out that both are simple multiwavelengths aspects of the same phenomena \citep{Tendulkar-2016}. In our case, the FRBs are triggered by the AQN, an external cause to the NS, which is not associated with SGRs, which are thought to be produced by rupture in the NS crust subsequently leading to flares \citep{Thompson-1995,Thompson-2002}.

%-------------------------------------------------------------------
\section*{Acknowledgements}\label{sec:Acknowledge}

This research was supported in part by the Natural Sciences and Engineering
Research Council of Canada. We would like to thank Emily Petroff for discussions about her FRB catalogue, and Jeremy Heyl, Maxim Lyutikov, Kiyoshi Masui for useful discussions.

%\medskip
%\newpage

%\bibliographystyle{mnras}
\bibliography{fast_radio_burst}

\appendix
\section{Magnetic Helicity ${\cal{H}}$ and its role in magnetic reconnection}\label{sect:helicity}

The main goal of this Appendix is to overview some important results on the magnetic helicity  which
is a topological invariant, and represents the observable which characterizes the dynamics of the magnetic reconnection. Needless to say that the magnetic reconnection,  according to our  proposal, plays the crucial role in transforming the static magnetic energy into the bursts which result in FRBs in the NS system, or  flares in case of the  solar  corona. The rate how the magnetic reconnection proceeds  is precisely related to the rate how the magnetic helicity varies and dissipates with time in a given system.  In other words,
the time scale of the dissipation of the magnetic helicity is directly related to the time scale of the magnetic  reconnection, and eventually to the time scales (\ref{duration}) for    $\tau_{\rm FRB}$ for the NS and duration of the   flares in case of the Sun.

We start with definition of the magnetic helicity in volume $V$ which can be represented as follows \cite{choudhury}
\be
\label{helicity}
{\cal{H}}\equiv \int_V \vec{A}\cdot \vec{{\cal{B}}} dV,
\ee
 where $\vec{A}$ is the vector potential corresponding to the magnetic field $\vec{{\cal{B}}}=\vec{\nabla}\times\vec{A}$.
 It is known that the magnetic helicity ${\cal{H}}$ in general is not a gauge invariant observable  because the gauge potential
$\vec{A}$ is not a gauge invariant object. However, if one  requires that the magnetic field is tangent on the surface boundary $\partial V$ of $V$, i.e. $\vec{{\cal{B}}}\cdot \vec{n}|_{\partial V}=0$, the magnetic helicity becomes well defined gauge invariant object, see e.g. \cite{choudhury}.

In simplest case when the magnetic configuration can  be represented in form of two interlinked  (but not overlapping) tubes
with fluxes $\Phi_1$ and $\Phi_2$, the magnetic helicity  ${\cal{H}}$ counts its linking number, i.e. ${\cal{H}}=2\Phi_1\Phi_2$
  is proportional to  an integer linking number if fluxes $\Phi_1$ and $\Phi_2$ are quantized. This is precisely the reason why
the magnetic helicity  is the topological invariant and cannot be easily changed during its evolution. In fact,
the crucial property of the  magnetic helicity  ${\cal{H}}$ is that it  is exactly  conserved  during the  time evolution in ideal
 MHD \cite{choudhury}. It is also known that the  magnetic helicity  ${\cal{H}}$  is odd under the $\cal{P}$   symmetry
 corresponding to : $\vec{x}\rightarrow -\vec{x}$ transformations. Furthermore, it is  known that in most astrophysical systems, including the NS,
 the  magnetic helicity  ${\cal{H}}$ is very large. This implies that the magnetic helicity can be only induced if there are $\cal{P}$  violating processes producing a large coherent effect on macroscopic scales.  We do not address in the present work the question on how the magnetic helicity  was generated in the first place in NS   referring to the  recent proposal \cite{Charbonneau:2009ax} for review and references. Instead, our goal here  is to study how the magnetic helicity  ${\cal{H}}$ evolves and dissipates as a result of the magnetic reconnection.

 The  magnetic helicity  ${\cal{H}}$  is  measured in $[{\rm Mx}^2]$, where $[{\rm Mx}]$  is a unit of magnetic flux in Gaussian units, i.e.  $[{\rm Mx}] =[{\rm G}\cdot {\rm cm}^2]$. In what follows we also need the expression for the temporal variation of magnetic helicity  as it is directly related to the dissipation rate. Differentiating of eq.  (\ref{helicity}) one arrives to
 \be
 \label{time-derivative}
 \frac{d{\cal{H}}}{dt}=-2\int_V \vec{E}\cdot \vec{{\cal{B}}} ~dV +\int_{\partial V}\vec{J}_{\cal{H}}\cdot d\vec{S}, ~~~~ \vec{J}_{\cal{H}}\equiv 2\vec{A}\times \vec{E}+\vec{A}\times \frac{\partial \vec{A}}{\partial t},
  \ee
  see e.g. \cite{Pariat2015},\cite{Yang2017} with explicit derivations. The key point for our discussions is that the volume integral precisely describes the dissipation of the magnetic helicity  ${\cal{H}}$ and it identically vanishes in ideal MHD where $\vec{E}=-\vec{v}\times \vec{{\cal{B}}}$. The second, surface term is irrelevant for our studies as it describes the transformation of the magnetic helicity from one form to another and is expressed in terms of the boundary terms. The expression for $\vec{J}_{\cal{H}}$ can be interpreted as the magnetic helicity current.

  The key observation for our studies is the fact that the dissipation term in (\ref{time-derivative}) is proportional to $ \sim \vec{E}\cdot \vec{{\cal{B}}}$ which is precisely the $E\&M$ configuration reconstructed from the requirement to match the FRB radiation during the burst due to the magnetic reconnection  as formula (\ref{E}) states. As explained in the text the induced electric field parallel to the original static magnetic field is absolutely required feature for the successful magnetic reconnection.

The only additional comment we would like to make here is that only a small amount\footnote{To be precise, ref.\cite{Yang2017} states that   8\%    of the total injected helicity $-8.5\cdot 10^{42} Mx^2$ will dissipate as a result of non-ideal MHD with fast magnetic reconnection. For comparison, according to \cite{Pariat2015} the dissipated helicity represents less than 3\% of the initial helicity.} of the total helicity dissipates during the reconnection.  It is in line with arguments of refs.  \cite{kumar_1, kumar_2} and with our estimates in Section \ref{FRB_parameters} that parameter $\beta_{\rm in}$ must be numerically small. In fact, one can argue that smallness of the  parameters $\beta_{\rm in}$ entering (\ref{duration}) and (\ref{E})
  reflects the main feature observed in numerical analysis of refs \cite{Pariat2015},\cite{Yang2017} that the dissipation is numerically suppressed and represents a few \% effect.

\end{document}